\documentclass[10pt, lettersize,journal]{IEEEtran}
\IEEEoverridecommandlockouts
\usepackage{ifpdf}
\usepackage{comment}
\usepackage{cuted}
\usepackage{cite}
\ifCLASSINFOpdf
\usepackage[pdftex]{graphicx}
\else
\usepackage[dvips]{graphicx}
\fi
\usepackage{mdwlist}
\usepackage{derivative}
\usepackage{commath, amsmath}
\usepackage{amsfonts}

\usepackage{stfloats}% <-- added
\usepackage{lipsum}
\usepackage{mathtools}
\usepackage{verbatim}
\usepackage{color,soul}
\usepackage{ntheorem}
\usepackage{algorithm}% http://ctan.org/pkg/algorithm
\usepackage{algpseudocode}% http://ctan.org/pkg/algorithmicx
\usepackage{array}
\usepackage[utf8]{inputenc}
\usepackage{tabularx,ragged2e,booktabs,caption}
\newcolumntype{C}[1]{>{\Centering}m{#1}}

\usepackage{fixltx2e}
\usepackage{stfloats}
\ifCLASSOPTIONcaptionsoff
\usepackage[nomarkers]{endfloat}
\let\MYoriglatexcaption\caption
\renewcommand{\caption}[2][\relax]{\MYoriglatexcaption[#2]{#2}}
\fi
\usepackage{url}
\usepackage{subcaption}
\usepackage[font={small}]{caption}
\usepackage[utf8]{inputenc}
\usepackage[english]{babel}
\usepackage{tabularx}
\usepackage{mathtools, nccmath}
\usepackage[utf8]{inputenc}
\usepackage[english]{babel}
\usepackage{ntheorem}
\theoremstyle{plain}
\theorembodyfont{}
\theoremsymbol{}
\theoremprework{}
\theorempostwork{}
\theoremseparator{.}

\newtheorem{remark}{Remark}
\newtheorem{lemma}{Lemma}
\usepackage{environ}         % provides \BODY
\usepackage{etoolbox}        % provides \ifdimcomp
\newlength{\myl}
\let\origequation=\equation
\let\origendequation=\endequation
\RenewEnviron{equation}{
\settowidth{\myl}{$\BODY$}                       % calculate width and save as \myl
\origequation
\ifdimcomp{\the\linewidth}{>}{\the\myl}
{\ensuremath{\BODY}}                             % True
{\resizebox{\linewidth}{!}{\ensuremath{\BODY}}}  % False
\origendequation
}

\author{Sheikh~Salman~Hassan,~\IEEEmembership{Student Member, IEEE,} Do Hyeon Kim, Yan Kyaw Tun,~\IEEEmembership{Member, IEEE}, Nguyen H. Tran,~\IEEEmembership{Senior Member, IEEE}, Walid Saad,~\IEEEmembership{Fellow, IEEE}, and~Choong~Seon~Hong,~\IEEEmembership{Senior Member, IEEE}
\thanks{Sheikh Salman Hassan, Do Hyeon Kim, Yan Kyaw Tun, and Choong Seon Hong  are with the Department of Computer Science and Engineering, Kyung Hee University, Yongin-si, Gyeonggi-do 17104, Rep. of Korea, e-mails:{\{salman0335,  doma, ykyawtun7, cshong\}@khu.ac.kr}.}
\thanks{Nguyen H. Tran is with the School of Computer Science, The University of Sydney, Sydney, NSW 2006, Australia (e-mail: nguyen.tran@sydney.edu.au).}
\thanks{Walid Saad is with the Bradley Department of Electrical and Computer Engineering, Virginia Tech, VA, 24061, USA, and the Department of Computer Science and Engineering, Kyung Hee University, Yongin-si, Gyeonggi-do 17104, Rep. of Korea, email:{\{walids@vt.edu\}}.}}
\begin{document}
\title{Seamless and Energy Efficient Maritime Coverage in Coordinated 6G Space-Air-Sea Non-Terrestrial Networks}
\maketitle
\begin{abstract}
Non-terrestrial networks (NTNs), which integrate space and aerial networks with terrestrial systems, are a key area in the emerging sixth-generation (6G) wireless networks. As part of 6G, NTNs must provide pervasive connectivity to a wide range of devices, including smartphones, vehicles, sensors, robots, and maritime users. However, due to the high mobility and deployment of NTNs, managing the space-air-sea (SAS) NTN resources, i.e., energy, power, and channel allocation, is a major challenge. The design of a SAS-NTN for energy-efficient resource allocation is investigated in this study. The goal is to maximize system energy efficiency (EE) by collaboratively optimizing user equipment (UE) association, power control, and unmanned aerial vehicle (UAV) deployment. Given the limited payloads of UAVs, this work focuses on minimizing the total energy cost of UAVs (trajectory and transmission) while meeting EE requirements. A mixed-integer nonlinear programming problem is proposed, followed by the development of an algorithm to decompose, and solve each problem distributedly. The binary (UE association) and continuous (power, deployment) variables are separated using the Bender decomposition (BD), and then the Dinkelbach algorithm (DA) is used to convert fractional programming into an equivalent solvable form in the subproblem. A standard optimization solver is utilized to deal with the complexity of the master problem for binary variables. The alternating direction method of multipliers (ADMM) algorithm is used to solve the subproblem for the continuous variables. Our proposed algorithm provides a suboptimal solution, and simulation results demonstrate that the proposed algorithm achieves better EE than baselines.
\end{abstract}
\IEEEpeerreviewmaketitle
\section{Introduction}
\IEEEPARstart{R}{}esearch on {6G} wireless networks is currently underway in both academia and industry \cite{walid6G}. One major component of 6G networks is non-terrestrial networks (NTNs) that consist of space and aerial-based networking \cite{6G_magazine}. NTNs are expected to provide global connectivity to regions and areas that are out of reach of existing terrestrial networks. For instance, NTNs can provide wireless network access to maritime users, called low-end user equipment (UE), that cannot directly connect to any satellite. In particular, these low-end UEs can get network services from an aerial access network near them \cite{dang2020should}. However, high-end UEs can directly connect with low-earth orbit (LEO) satellites. Thus, a coordinated space-air-sea (SAS)-based NTN network can extend the existing coastline base stations (CBSs) coverage seamlessly. This heterogeneous SAS-NTN can meet the increasing maritime network requirements, i.e., seamless, energy-efficient, and high throughput coverage. The design of NTNs faces many challenges, including the coordinated integration of space, air, and sea platforms. While some of these issues have been addressed in the past (see Section \ref{relwork}), nonetheless, the joint maritime users' fronthaul and backhaul communication mechanisms and UAV deployment techniques in heterogeneous networks are missing.

The main contribution of this paper is a novel SAS-NTNs architecture that enabled each maritime UE to connect with the terrestrial networks. For reliable communication in this network, maritime users with a high gain antenna, i.e., high-end UEs (HUEs), can directly associate with a LEO satellite or a CBS depending on their vicinity. However, low-end UEs (LUEs) cannot effectively communicate with a LEO satellite or a CBS due to low antenna gain \cite{hybridnetwork}. In particular, LUEs require assistance from UAVs, i.e., UAVs could transmit LUEs' data to a LEO satellite or CBS, using aerial-to-satellite (A2S) links \cite{S2S} or aerial-to-ground (A2G) links. These LUEs limit themselves for long transmission distance due to power consumption constraints  \cite{multiantenna}. As a result, the UAV is regarded as an effective mode of communication for LUEs in the maritime environment. UAVs are quickly deployed on the place of interest, which is critical in isolated maritime regions \cite{iotsurvey}. Our key contributions are summarized as follows:
\begin{itemize}
\item We propose a novel heterogeneous SAS network architecture for next-generation maritime mobile networks. To serve maritime users, we propose the use of a LEO satellite coupled with UAVs and CBSs for the service provisioning of low-end and high-end UEs. 
	    
\item We study the problem of resource management in the SAS-NTNs to optimize resource block allocation, transmit power control, and UAVs deployment for maximizing network energy efficiency (EE). An energy efficiency maximization problem is formulated by considering the constraint of the limited payload of UAVs and also their power consumption.

\item The problem of resource allocation in the SAS maritime network is formulated as a mixed-integer nonlinear programming (MINLP) problem. The goal is to optimize the utility function considering the energy efficiency of the network.
        
\item Due to the problem's high complexity, we propose a novel algorithm to solve the MINLP problem, composed of the Bender's decomposition (BD), Dinkelbach algorithm (DA), alternating direction method of multipliers (ADMM) algorithm, and an optimization solver.
        
\item The BD algorithm decomposes the main problem into a master problem and another subproblem to obtain the solution efficiently. The variables of the original problem are divided into two subsets so that a first-stage master problem is solved over the first set of variables, and the values for the second set of variables are determined in a second-stage subproblem for a given first-stage solution.
        
\item We use the Dinkelbach algorithm for the subproblem to transform fractional programming into an equivalent form and adopt ADMM in the inner loop to distributedly solve the continuous large-scale problem. We use the optimization solver in the master problem to solve pure integer programming with complexity reduction considerations.

\item We evaluate the performance of our proposed algorithm in the simulation. Our numerical results demonstrate that the proposed algorithm achieves a near-optimal solution and outperforms the other baselines. The proposed algorithm achieves EE up to $9\%$ and 10$\%$ compared to greedy and dynamic algorithms, respectively.
\end{itemize}

The rest of this paper is organized as follows. In Section \ref{relwork}, the research background and the objective are presented. Section \ref{sysmodel} represents the system model. In Section \ref{form}, we formulate the optimization problem. In Section \ref{decom}, the problem decomposition and the proposed algorithms are presented. Numerical results and corresponding analyses are provided in Section \ref{sim}. The main notations are given in Table \ref{notations}.\\

\section{Related Work}
\label{relwork}
We now review the prior works in the area of NTNs, satellite and UAV-based networking, maritime communication, and their combinations. We particularly show the classification of maritime users, i.e., we can provide networking resources to each maritime user based on their antenna gain and feasible connectivity. Despite significant advances, prior works remain limited as they do not address the challenges of maritime users' resource allocation based on their classification and overall network energy efficiency by jointly considering all the network nodes involved in SAS-NTNs.

Various elements of NTNs such as LEO satellite constellation deployment have been examined in the literature, including satellite number minimization \cite{howmanysat} and \cite{howmanysat2}, coverage maximization \cite{sat_const_1}, communication latency reduction \cite{sat_delay_1}, and heterogeneous network design \cite{sat_iot_1}. For satellite constellation optimization, several intelligence algorithms are used, including the genetic algorithm (GA), differential evolution (DE), immune algorithm, and particle swarm optimization (PSO) [24]. The work in \cite{howmanysat2} developed a non-dominated sorting evolutionary algorithm for regional LEO satellite constellation design to match UE needs while reducing satellite cost. The authors in \cite{meziane2016optimization} proposed a satellite constellation for continuous mutual regional coverage based on the evolutionary optimization approach. It has been explored the relationship between the coverage ratio and the number of satellites. The works in \cite{sat_const_1} and \cite{savitri2017satellite} used an evolutionary algorithm to optimize the coverage of target areas while designing regional satellite constellations. To reduce the end-to-end latency, authors in \cite{sat_delay_1} devised a progressive satellite constellation network building method. The work in \cite{leo_iot_2} investigated the use of LEO satellites within the context of the Internet of Things. The performance of satellite constellation design with a few intelligent algorithms, i.e., GA, DE, immunity algorithm, and PSO, was compared in \cite{intel_algo_sat} to enhance satellite coverage capabilities. NTNs face a slew of new difficulties, including high bit error rates, extended propagation delays, and unreliable connections. As a result, it's important to think about how to incorporate network operations into NTNs efficiently. 

A significant number of related prior works on NTNs focused on solutions that can improve the connectivity of ground networks by using UAVs \cite{UAV_only_1, UAV_only_2, UAV_only_3, UAV_only_4, UAV_only_5, UAV_only_6, UAV_only_7, UAV_only_8,UAV_only_9,UAV_only_10, UAV_only_11, UAV_only_12, UAV_only_13, UAV_only_14,UAV_only_15,UAV_only_16}. The authors in \cite{UAV_only_1, UAV_only_2, UAV_only_3, UAV_only_4, UAV_only_5, UAV_only_6, UAV_only_7,UAV_only_8,UAV_only_9,UAV_only_10, UAV_only_11, UAV_only_12, UAV_only_13} concentrate primarily on static type UEs. The authors in \cite{UAV_only_14,UAV_only_15,UAV_only_16} investigated how to optimize the ergodic achievable rate by remotely monitoring the UAV trajectory to deal with the moving UEs. Meanwhile, the rotary-wing UAV placement problem is widely studied to provide useful results. However, in the case of fixed wing UAVs, the key issue is their optimum transmission and trajectory. In particular, the trajectory of UAVs is determined by taking into account the maximum velocity or acceleration to achieve the maximum sum rate, the minimum service flight time, and the optimum energy efficiency in the network. 

There have been a number of recent works that looked at the co-existence of UAVs and ground base stations (GBSs) \cite{Coexistence_UAV_TBS_1, Coexistence_UAV_TBS_2, Coexistence_UAV_TBS_3, Coexistence_UAV_TBS_4, Coexistence_UAV_TBS_5}. The use of a GBS as a central controller for a UAV-based network was proposed in \cite{Coexistence_UAV_TBS_1} to maximize the sum rate by taking radio access and backhaul links into account. To counter the dynamics of UAV-based networks, the authors in \cite{Coexistence_UAV_TBS_2} proposed the idea of multihop backhaul networks. The works in \cite{Coexistence_UAV_TBS_3, Coexistence_UAV_TBS_4, Coexistence_UAV_TBS_5} analyzed the outage probability of the GBS and UAV networks. In \cite{Coexistence_UAV_TBS_5}, the authors studied the sum rate of the network by taking outage probability into account. The authors in \cite{maritime_1} studied a GBS and multiple offshore relay nodes for a cooperative multicast communication strategy for maritime users based on combined beamforming (BF) optimization and relay design. The authors in \cite{maritime_2} provide a maritime communication network design in which a GBS provides wireless backhaul for shipborne base stations, while the shipborne base stations act as mobile access points for user ships Although GBS may provide end-users with real-time services and high data rates, their network coverage in marine communication is restricted. Furthermore, deploying expensive floating edge computer equipment in deep oceans is too expensive. Consider UAV technology, which necessitates the use of edge servers to deliver seamless and real-time services to moving boats. An edge server's coverage diameter (e.g., a tiny cell base station) is often less than 300 m. As a result, moving vessels will encounter frequent handovers in GBS networks. More critically, vessels engaged in marine communication may lose network connectivity.

In addition, several recent works \cite{Coexistence_UAV_Satellite_1, Coexistence_UAV_Satellite_2, Coexistence_UAV_Satellite_3, Coexistence_UAV_Satellite_4, Coexistence_UAV_Satellite_5, Coexistence_UAV_Satellite_6} studied the use of multi-layer heterogeneous network architectures for NTNs. Specifically, in \cite{Coexistence_UAV_Satellite_3}, the authors studied the problem of UAV satellite integration for a hybrid flying autonomous vehicle. Meanwhile, the authors in \cite{Coexistence_UAV_Satellite_4} investigated the optimal altitude of UAV to analyze the suitable coordination in case of communication between satellite and UAV with a focus on reducing latency. Similarly, the work in \cite{Coexistence_UAV_Satellite_5} analyzed the coverage and rate of a multi-UAV network in a disaster scenario. In \cite{Coexistence_UAV_Satellite_6}, the authors considered an airborne cellular network and studied the problem of resource allocation, i.e., transmit power control for the various time-critical application.

In this prior art \cite{Coexistence_UAV_TBS_1, Coexistence_UAV_TBS_2, Coexistence_UAV_TBS_3, Coexistence_UAV_TBS_4, Coexistence_UAV_TBS_5, Coexistence_UAV_Satellite_1, Coexistence_UAV_Satellite_2, Coexistence_UAV_Satellite_3, Coexistence_UAV_Satellite_4, Coexistence_UAV_Satellite_5, Coexistence_UAV_Satellite_6}, the spectrum resources that have a direct effect on the EE of a SAS-NTN were not taken into account. To examine the relationship between the satellite backhaul, the CBS backhaul links, and the radio access linkages in SAS networks, it is essential to consider the joint problem of user association, resource allocation across all communication links within the SAS network, and the deployment of UAVs above sea region which is missing in the literature. However, the preceding studies all regarded satellites, UAVs, and CBS to be the only network node of SAS-NTNs. Furthermore, prior works primarily consider direct connections, leaving out backhaul transmission. In response to the aforementioned finding, we offer a unique SAS-NTNs architecture to overcome the maritime UEs communication problem according to their classification. For SAS-NTNs EE, a combined problem of UE association, power control, and UAV deployment is developed.\\

\section{System Model}
\label{sysmodel}
\begin{figure}[t] 
\centering 
\includegraphics[width=\columnwidth]{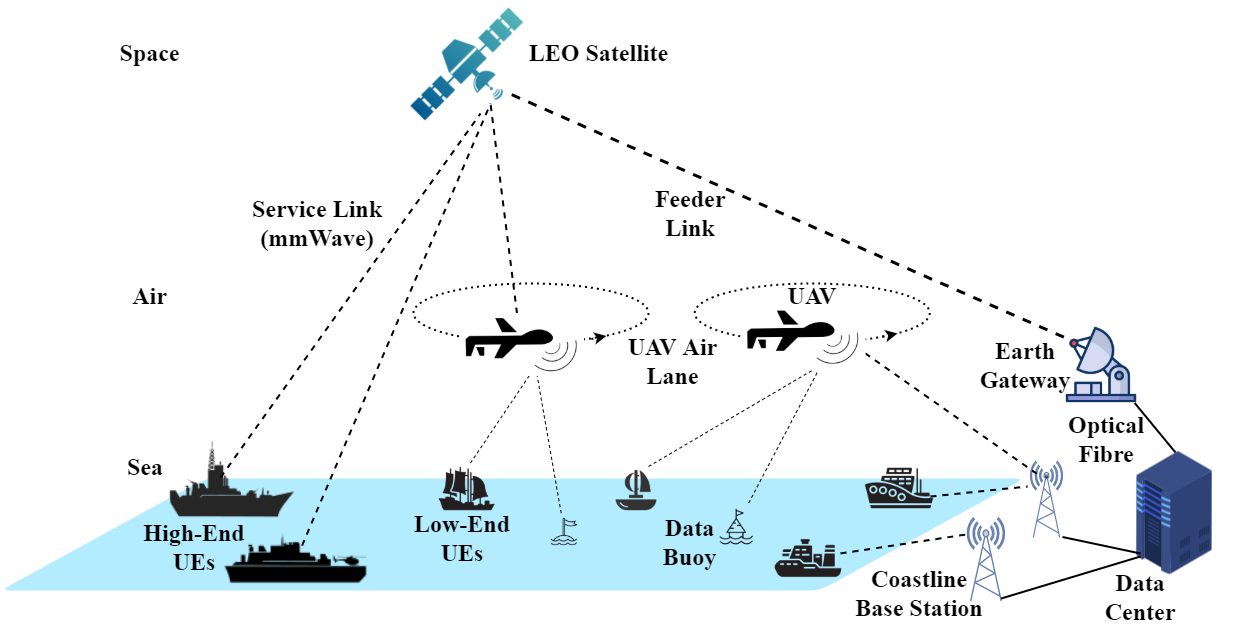}
\caption{Illustration of space-air-sea networks architecture.}
\label{fig:sysmodel}
\vspace{-0.1in}
\end{figure}

As shown in Fig. \ref{fig:sysmodel}, we consider a realistic heterogeneous SAS maritime communication network consisting of a LEO satellite\footnote{Hereinafter, the satellite is considered as a LEO satellite unless otherwise stated.} $s$, a set $\mathcal{U}$ of $U$ UAVs that serve as aerial base stations (ABSs)\footnote{Hereinafter, the UAV is considered as an ABS unless otherwise stated.}, and a set $\mathcal{C}$ of $C$ CBSs. The coverage area of each CBS is planar in the sea with a radius $\nu$ centered at $(0, 0) \in R^2$. We define a set $\mathcal{M}_l$ of $M_l$ LUEs and a set $\mathcal{M}_h$ of $M_h$ HUEs. We define a set $\mathcal{M} = \mathcal{M}_l \cup \mathcal{M}_h$ of $M$ maritime UEs. To capture the dynamic nature of the nodes, i.e., satellite, ABSs, and UEs. We consider the network within a certain time duration $T$ that is divided into a set $\mathcal{N}$ of $N-1$ time slots. Due to the short duration in each time slot $n$, the network configuration is considered fixed. Therefore, we will then analyze network performance in a one-time slot. CBS can serve the coastal region, but its broadband services are limited due to significant non-line-of-sight path loss. Various UEs are present in the waters, such as cruise ships and vessels equipped with high gain antennas that can directly connect with the satellite or a CBS depending upon their location. Conversely, LUEs, e.g., seamen, fishers, offshore platform users, maritime internet-of-things (MIoT) devices within the coverage region of the satellite or CBSs, cannot directly access their services and require service by ABSs. However, the satellite and CBSs will provide radio access to the HUEs and backhaul services to the deployed ABSs. \\

\subsection{Communication Model}
In the considered scenario, all the communication links operate over the Ka-band (26.5 – 40 GHz), which is a well-defined millimeter wave (mmW) range suitable for satellite communication and future 5G links, as discussed in \cite{wwWave_walid}. Directional transmissions over the mmW band are unavoidable to resolve the incredibly high path loss. Consequently, in compliance with established standards such as IEEE 802.15.3c \cite{mmwave}, a service provider node uses the multi-access time division scheme (TDMA) to provide services to its maritime users. Each maritime UE is a set element that seeks services that must be scheduled over the mmW band at each time slot $n$. In practice, the mmW transceiver must align its beams during a beam training stage so that the maximum beamforming gain is achieved. This phase of training will introduce a non-negligible TDMA system overhead, which can be particularly important as the number of mmW users increases. We assume that a beam training overhead time per transmission over the mmW band had already been established for the training phase as given in \cite{mmwave}. Moreover, due to the considerable distance between the satellite and LUEs, the interference experienced by LUEs from the satellite is negligible. Although each ABS shares the same frequency spectrum to provide downlink services to LUEs, therefore these LUEs experience interference from non-associated ABSs. Similarly, the satellite and CBSs share the same frequency spectrum, which also leads to interference at the ABSs and HUEs.\\

\subsection{Network Deployment Model}
The satellite orbits at an altitude $h_s$ (from the sea surface), and it provides wireless backhaul connectivity to ABSs and radio access to HUEs in its coverage region. The ABSs are sparsely deployed on the sea near coastal areas \cite{Sea_routes} to provide connectivity to the set of LUEs in their coverage region. This model considers that each ABS has a total mission flight time $T$. During $T$, each ABS $u$ must serve LUEs in its coverage region. As mentioned earlier, the UAV flight time $T$ is discretized into a set $\mathcal{N}$ of $N$ equally-spaced time slots with the length of each time slot is given as $L_u = \frac{T}{N}$. Moreover, the value of $N$ should be properly chosen to guarantee that the ABS location remains unchanged within each time slot and fulfill the network requirements, i.e., task processing. Each ABS $u$ flies at a fixed height $h_u$ above the sea surface in each time slot $n$. Thus, at each slot $n$, the position of each ABS $u$ in three-dimensional coordinates will be:
\begin{equation}
\boldsymbol{d}_{u}(n)=\left[\left(x_{u}(n), y_{u}(n), h_{u}\right)\right]^{T}, \forall u \in \mathcal{U}, \forall n \in \mathcal{N}.
\end{equation}
Similarly, the position of an LUE $m_l$ is $\boldsymbol{d}_{m_l} = (x_{m_l}(n), y_{m_l}(n), 0)$, and the position of a HUE $m_h$ will be $\boldsymbol{d}_{m_h}  = (x_{m_h}(n), y_{m_h}(n), 0)$. Both types of UEs will be distributed in a specified region at sea. Moreover, the position of each CBS can be represented by $\boldsymbol{d}_{c}  = (x_{c}, y_{c}, 0)$. Additionally, the satellite position can be given by $\boldsymbol{d}_{s}  = (x_{s}, y_{s}, h_s)$, which remains constant during the studied time. The ABS backhaul service can be provided by the satellite, or a CBS \cite{HYBRID_NET} depends upon its position in each time slot $n$.\\

\subsection{ABS Energy Consumption Model}
The ABS's overall energy consumption is made up of two parts. The first is an energy associated with communications, and this energy is generated by radiation, signal processing, and other electronics. The other component of energy is propulsion, which is required to keep the UAV aloft as well as to support its movement if necessary. We consider an autonomous ABS that can operate as an aerial relay node and a base station in a sea environment. This ABS can perform resource allocation, dynamic mission planning, inter-cell handover, and other tasks \cite{ee_net}. The maximum speed of the ABS in each time slot is $v_{\textrm{max}}$. Thus, the maximum distance that an ABS travels within each time slot will be $L_u v_{\textrm{max}}$. The energy consumption of the ABS for traveling from one location to another in each time slot can be given by \cite{UAV_Energy_model_Rui}:
\begin{equation}
\label{UAV_Energy_model}
E_{u}^{\text {flight }}(n)=\left(\kappa_u\left\|v_{u}(n)\right\|^{3}+\frac{\zeta_u}{\| v_{u}(n)\|}+\frac{\zeta_u\|\mu_{u}(n)\|^{2}}{q^{2} \| v_{u}(n)\|}\right)+\frac{\Delta j}{\Lambda}, \forall u \in \mathcal{U},
\end{equation}
where $\kappa_u$ and $\zeta_u$ are constants which depends on the ABS characteristics (e.g., weight, wing size, air density, etc.), $\Lambda$ is any infinitesimal time step, $q$ is the gravitational acceleration, and $\Delta J = \frac{1}{2} \pi \left( \| v_n(n+1) \|^2 - \|v_n(n)\|^2 \right)$ is the kinetic energy. Here, $\pi$ is the mass of the UAV's payload\footnote{Hereinafter, we ignore the change in weight of the ABS as more battery and fuel are consumed over time for simplicity.}, $v_u$ is the speed, and $\mu_u$ is the acceleration of each ABS $n$. We assume that the distance of each ABS to UEs, CBSs, and the satellite remains constant within each time slot $n$. The operating time of each ABS $u$ is calculated primarily by the fuel for flying and the battery for communication. The fuel of a fixed-wing ABS is assumed to be large enough for our studied time of the network performance. To validate the obtained energy consumption model, we investigate the case of steady, straight and fixed heights with constant speed $V$, i.e., $v(n)$ = $V$ and $\mu_u(n)$ = $0$. Then, (\ref{UAV_Energy_model}) can be modified as:
\begin{equation}
\label{simple_energy_UAV_model}
E_{u}^{\text {flight }}(n) = \big( \kappa_uV^3 + \frac{\zeta_u}{V} \big).
\end{equation}
Equation (\ref{simple_energy_UAV_model}) is a classical model of the energy consumption in aerodynamics \cite{thermodynamics}. The model comprises two components in (\ref{simple_energy_UAV_model}), where $V^3$ is used to overcome the parasite drag and $\frac{1}{V}$ allows overcoming the elevated drag. Therefore, the flying ABSs power can be calculated as:
\begin{equation}
P_u^{\mathrm{flight}}(n) = E_{u}^{\text {flight }}(n) \times L_u,
\end{equation}
where $L_u$ represents the duration of each time step.\\
\setlength{\arrayrulewidth}{0.10mm}
\setlength{\tabcolsep}{3.5pt}
\renewcommand{\arraystretch}{1}
\begin{table}[t]
\centering
\captionof{table}{SUMMARY OF NOTATIONS} 
\label{notations} 
\begin{tabular}{|l|l|}
\hline
\textbf{Notation} & \textbf  {Description}\\ \hline \hline
$\boldsymbol{d}_{u}$             & 3D coordinates of ABS $u$ $\rightarrow$ $d_{u} = \{x_{u}, y_{u}, h_{u}\}$\\ \hline
$\boldsymbol{d}_{m_l}$           & Position of low-end UEs $m_l$ $\rightarrow$ $d_{m_l} = \{x_{m_l}, y_{m_l}, 0\}$\\ \hline
$\boldsymbol{d}_{m_h}$           & Position of high-end UEs $m_h$ $\rightarrow$ $d_{m_h} = \{x_{m_h}, y_{m_h}(n), 0\}$\\ \hline
$\boldsymbol{d}_{s}(n)$         & Position of the satellite $s$  $\rightarrow$ $d_{s} = \{x_{s}, y_{s}, h_s\}$\\ \hline
$\boldsymbol{d}_{c}$             & Position of the CBS $c$ $\rightarrow$ $d_{c} = \{x_{c}, y_{c}, 0\}$ \\ \hline
$K_u$                           & Set of resource blocks allocated for each ABS $u$\\ \hline
$Z_s$                           & Set of resource blocks allocated for a LEO satellite\\ \hline
$Y_c$                           & Set of resource blocks allocated for each CBS\\ \hline
$g_{u,{m_l}}$            		& Radio-access channel gain from ABS $u$ to low-end UE $m_l$\\ \hline % for the $k$-th resource block
$g_{s,u}$           		    & Backhaul channel gain from satellite $s$ to ABS $u$\\ \hline % by $z$-th resource block\\ \hline
$g_{c,u}$						& Backhaul channel gain from CBS $c$ to ABS $u$\\ \hline % by resource block $y$\\ \hline
$g_{s,{m_h}}$          			& Radio-access channel gain from satellite $s$ to high-end UE $m_h$\\ \hline % by resource block $z$\\ \hline
$g_{c,{m_h}}$ 					& Radio-access channel gain from CBS $c$ to high-end UE $m_h$\\ \hline % by resource block $y$\\ \hline
$p_{u, {m_l}}$					& Transmit power from ABS $u$ to low-end UE $m_l$\\ \hline % over the resource block $k$\\ \hline
$p_{s,u}$ 						& Transmit power from satellite $s$ to ABS $u$\\ \hline % over the resource block $z$\\ \hline
$p_{c,u}$						& Transmit power from UAV $u$ relay node to CBS $c$\\ \hline % in the resource block $y$\\ \hline
$p_{s, {m_h}}$					& Transmit power from satellite $s$ to ABS $u$\\ \hline % over the allocated resource block $z$\\ \hline
$p_{c, {m_h}}$					& Transmit power from CBS $c$ to ABS $u$\\ \hline % over the allocated resource block $y$\\ \hline
$\gamma_{u,{m_l}}$ 				& SNR between ABS $u$ and low-end UE $m_l$\\ \hline
$\gamma_{s,u}$					& SNR between satellite $s$ and ABS $u$\\ \hline
$\gamma_{c,u}$      			& SNR between CBS $c$ and ABS $u$\\ \hline
$\gamma_{s,{m_h}}$     			& SNR between satellite $s$ and high-end UE $m_h$\\ \hline
$\gamma_{c,m_h}$     			& SNR between CBS $c$ and high-end UE $m_h$\\ \hline
$r_{{m_l},u}$                   & Achievable datarate from ABS $u$ to low-end UE $m_l$\\ \hline
$r_{s,u}$                       & Achievable datarate from satellite $s$ to ABS $u$\\ \hline
$r_{c,u}$                       & Achievable datarate from CBS $c$ to ABS $u$\\ \hline
$r_{s,m_h}$                     & Achievable datarate from satellite $s$ to high-end UEs $m_h$\\ \hline
$r_{c,m_h}$                     & Achievable datarate from CBS $c$ to high-end UEs $m_h$\\ \hline
\end{tabular}
\end{table}

\subsection{Low-End UE-ABS Data Link Analysis}
In the SAS network, each ABS $u$ is placed at a high-enough altitude to enable LoS transmission. Therefore, we use a general composite channel model coefficient that consists of both small-scale and large-scale fading between each ABS $u$ and the low-end UE $m_l$ at each time slot $n$, as follows:
\begin{equation}
g_{u,m_l}(n) =  \beta_{u,{m_l}}(n) \sqrt{\xi_{u,ml}(n)},
\end{equation}
where $\beta_{u,{m_l}}(n)$ is the small-scale fading coefficient with ${\mathbb E}[|\beta_{u,{m_l}}|^2]=1.53$ \cite{ursula} and $\xi_{u,ml}(n)$ is the large-scale fading coefficient. Each ABS knows the coordinates of LUEs and channel distribution information, i,e., $\xi_o$ and $|\beta_{u,{m_l}}|^2$. The large scale fading at each time slot $n$ will be:
\begin{equation}
\label{LUE_pathloss}
\xi_{u,ml}(n) = \frac{\xi_0}{\|\left( \boldsymbol{d}_{u}(n)-\boldsymbol{d}_{m_{l}}(n) \right) \|^2  }, \forall u\in\mathcal{U}, m_l\in\mathcal{M}_l, n\in\mathcal{N}.
\end{equation}
In (\ref{LUE_pathloss}), $\xi_0$ is the reference channel gain at 1m and $\|\left( \boldsymbol{d}_{u}-\boldsymbol{d}_{m_{l}} \right) \|^2 $ is 3D Euclidean distance between ABS $u$ and low-end UE $m_l$. We consider a  Rician distribution for modeling the small-scale fading between LUE $m_l$ and ABS $u$ to compensate for the LoS and multipath scatterers that can be experienced by each receiving LUE in the network. Specifically, adopting the Rician channel model is justified by the fact that the channel between ABS $u$ and LUE $m_l$ is primarily dominated by LoS \cite{khwaja}. Moreover, the Doppler effect due to mobility in network nodes is compensated by existing frequency synchronization techniques, i.e., phase-locked loop as discussed in \cite{phasedloop, phasedloop1, phasedloop2}. Each ABS shares the same set of resource blocks to provide downlink services to LUEs. Therefore, the interference in ABS-LUE link from non-associated ABSs and CBSs at time slot $n$ will be $ \Omega_{u,m_l} = \sum\limits_{\forall u' \ne u }\sum\limits_{\forall m_l' \ne m_l } p_{u'} g_{u', m_l'} + \sum\limits_{\forall c \in \mathcal{C} } p_c g_{c, m_l}$. Here $u'$ is non-associated ABSs, $p_{u'}$ is the transmit power of non-associate ABSs, and $p_c$ is the transmit power of CBSs. Thus, the signal-to-noise ratio (SINR) between this link can be given as:
\begin{equation}
\gamma_{u,{m_l}}(n) = \frac{p_{u,{m_l}}(n)  g_{u,{m_l}}(n)}{\Omega_{u,m_l} + \sigma^{2}},  \quad \forall u \in \mathcal{U} , m_l \in \mathcal{M}_l, n \in \mathcal{N},
\end{equation}
where $p_{u,{m_l}}(n)$ is the transmit power of ABS $u$ to low-end UE $m_l$ in the $k$th RB, and $\sigma^2$ is the the noise power. Moreover, following \cite{Zhu_IoT_LEO}, even when there is additional interference at the receiver, we suppose that the aggregate interference follows a Gaussian distribution and the corresponding power is incorporated into the noise term $\sigma^2$. The achievable data rate without transmission diversity between low-end UE $m_l$ and ABS $u$ in each time slot $n$ will be:
\begin{equation}
r_{u,{m_l}}(n) = B_u \log_2 \left(1 +  \gamma_{u,{m_l}}(n)  \right),
\end{equation}
where $B_u$ is the bandwidth of each RB $k$ over the band allocated from ABS $u$ to LUEs $m_l$ at time slot $n$.\\

\subsection{Satellite based ABS Backhaul Link Analysis}
The satellite provides backhaul services to the ABSs outside of the CBS coverage region. Therefore, we consider that the ABS $u$ and the satellite $s$ are equipped with one antenna each. Therefore, the channel model between ABS $u$ and the satellite $s$ can be define as:
\begin{equation}
g_{s,u}(n) = \beta_{s,u} (\xi_{s, u})^{-1/2},\quad \forall u \in \mathcal{U}, n \in \mathcal{N}.
\end{equation}
where $\beta_{s,u}$ is the Rician fading channel coefficient and $\xi_{s, u}$ represents large-scale fading for pathloss. The $d_{s,u} = \left( \sqrt{ (x_s - x_u)^{2} + (y_s - y_u)^{2} + (z_s - z_u)^{2} } \right)$ denotes the distance between satellite $s$ and ABS $u$. Thus, large-scale path loss on the mmW links will be given by \cite{mmwave_largescale}:
\begin{equation}
\xi_{s, u}(\mathrm{dB}) = \omega_{s, u} + \zeta_{s,u} 10 \log_{10} \Big(\frac{d_{s,u}(n)}{d_0}\Big)  + \psi_{s,u}
\end{equation}
where $\zeta_{s, u}$ is the slope of the fit (path loss exponent), $\omega_{s, u}$ indicate the intercept parameter (path loss at reference distance $d_0$) \cite{mmwave}, and $\psi_{s,u}$ models the deviation in fitting (dB) which is a zero mean Gaussian random variable with standard deviation $\delta_u$. The small scale fading coefficient will be:
\begin{equation}
\beta_{s,u} = \sqrt{\frac{K_{s,u}}{1+K_{s,u}}}+\sqrt{\frac{1}{1+K_{s,u}}}\Xi_{s,u},
\end{equation}
where $K_{s_u}$ is the Rician factor and $\Xi_{s,u} \sim \mathcal{N}(0,1)$. We can now simplify the channel gain:
\begin{equation}
g_{s,u}(n) = \Big( \frac{d_0}{d_{s,u}(n)} \Big)^{\frac{\zeta_{s,u}}{2}} 10^{-\frac{\omega_{s, u}+\psi_{s,u}}{20}}\Bigg( \sqrt{\frac{K_{s,u}}{1+K_{s,u}}}+\sqrt{\frac{1}{1+K_{s,u}}}\Xi_{s,u} \Bigg)
\end{equation}
The interference at time slot $n$ in this link from CBSs will be $\Omega_{s,u} = \sum\limits_{\forall c \in \mathcal{C}} p_c g_{c, u}$. Here $p_c$ is the transmit power of CBSs. Thus, the SINR between this link  can be given as:
\begin{equation}
\gamma_{s,u}(n) = \frac{p_{s,u}(n)  g_{s,u}(n)}{\Omega_{s, u}+\sigma^{2}},\quad \forall u \in \mathcal{U}, \forall n \in \mathcal{N},
\end{equation}
where $p_{s,u}(n)$ is the transmit of satellite $s$ in the $z$th RB to ABS $u$ at time slot $n$. The achievable data rate between satellite $s$ and ABS $u$ in each time slot $n$ can be calculated by Shannon capacity:
\begin{equation}
r_{s,u}(n) = B_s\log_2 \left(1 + \gamma_{s,u}(n) \right),
\end{equation}
where $B_s$ denotes the bandwidth allocated to the channel from satellite $s$ to UAV $u$ at time slot $n$.\\

\subsection{CBS based ABS Backhaul Link Analysis}
The CBS provides backhaul services to the ABSs near coastline under their coverage region. Therefore, the channel model between ABS $u$ and the CBS $c$ will be:
\begin{equation}
g_{c,u}(n) = \beta_{c,u} (\xi_{c, u})^{-1/2},\quad \forall c \in \mathcal{C}, u \in \mathcal{U}, n \in \mathcal{N},
\end{equation}
where $\beta_{c,u}$ is the Rician fading channel coefficient and $\xi_{c, u}$ represents the large-scale fading. $d_{c,u} = \left( \sqrt{ (x_c - x_u)^{2} + (y_c - y_u)^{2} + (z_c - z_u)^{2} } \right)$ is the distance between CBS $c$ and ABS $u$. Thus, large-scale path loss on the mmW links will be:
\begin{equation}
\xi_{c, u}(\mathrm{dB}) = \omega_{c, u} + \zeta_{c,u} 10 \log_{10} \Big(\frac{d_{c,u}(n)}{d_0}\Big)  + \psi_{c,u},
\end{equation}
where $\zeta_{c, u}$ is the slope of the fit (path loss exponent), $\omega_{c, u}$ is the intercept parameter (path loss at reference distance $d_0$) \cite{mmwave}, and $\psi_{c,u}$ models the deviation in fitting (dB) which is a zero mean Gaussian random variable with standard deviation $\delta_c$. The small scale fading coefficient will be:
\begin{equation}
\beta_{c,u} = \sqrt{\frac{K_{c,u}}{1+K_{c,u}}}+\sqrt{\frac{1}{1+K_{c,u}}}\Xi_{c,u},
\end{equation}
where $K_{c_u}$ is the Rician factor and $\Xi_{c,u} \sim \mathcal{N}(0,1)$. We can then simplify the channel gain:
\begin{equation}
g_{c,u}(n) = \Big( \frac{d_0}{d_{c,u}(n)} \Big)^{\frac{\zeta_{c,u}}{2}} 10^{-\frac{\omega_{c, u}+\psi_{c,u}}{20}}\Bigg( \sqrt{\frac{K_{c,u}}{1+K_{c,u}}}+\sqrt{\frac{1}{1+K_{c,u}}}\Xi_{c,u} \Bigg).
\end{equation}
The interference at time slot $n$ in this link from non-associated CBSs will be $ \Omega_{c,u} = \sum\limits_{\forall c' \ne c } p_{c'} g_{c', u} $. Here $c'$ is non-associated CBSs, and $p_{c'}$ is the transmit power of non-associated CBSs. Thus, the SINR between this link can be given as:
\begin{equation}
\gamma_{c,u}(n) = \frac{p_{c,u}(n)  g_{c,u}(n)}{\Omega_{c,u}+\sigma^{2}}, \forall c \in \mathcal{C}, \forall u \in \mathcal{U}, \forall n \in \mathcal{N},
\end{equation}
where $p_{c,u}(n)$ is the transmit of CBS $c$ to ABS $u$ over the $y$th RB at time slot $n$. The achievable data rate between CBS $c$ and ABS $u$ in each time slot $n$ can be calculated by Shannon capacity:
\begin{equation}
r_{c,u}(n) = B_c\log_2 \left( 1 + \gamma_{c,u}(n)  \right),
\end{equation}
where $B_c$ is the bandwidth allocated to the channel from CBS $c$ to ABS $u$ at time slot $n$.\\

\subsection{Satellite-HUEs Data Link Analysis}
In the SAS network, HUEs are considered with a high gain antenna that can directly connect with the satellite $s$. The satellite provides backhaul services to HUEs out of the CBS coverage region. Therefore, we consider that the satellite $s$ and the HUE $m_h$ are equipped with one antenna. Therefore, the channel model between the satellite $s$ and HUE $m_h$ will be:
\begin{equation}
g_{s,{m_h}}(n) = \beta_{s,{m_h}} (\xi_{s, {m_h}})^{-1/2},\quad \forall m_h \in \mathcal{M}_h, \forall n \in \mathcal{N},
\end{equation}
where $\beta_{s,{m_h}}$ is the Rician fading channel coefficient and $\xi_{s, {m_h}}$ represents the large-scale fading for pathloss. $d_{s,u} = \left( \sqrt{ (x_s - x_u)^{2} + (y_s - y_u)^{2} + (z_s - z_u)^{2} } \right)$ is the distance between satellite $s$ and HUE $m_h$. Thus, the large-scale path loss on the mmW link will be \cite{mmwave_largescale}:
\begin{equation}
\xi_{s, {m_h}}(\mathrm{dB}) = \omega_{s, {m_h}} + \zeta_{s,{m_h}} 10 \log_{10} \Big(\frac{d_{s,{m_h}}(n)}{d_0}\Big)  + \psi_{s,{m_h}}
\end{equation}
where $\zeta_{s, {m_h}}$ is the slope of the fit (path loss exponent), $\omega_{s, {m_h}}$ indicate the intercept parameter (path loss at reference distance $d_0$) \cite{mmwave}, and $\psi_{s,{m_h}}$ models the deviation in fitting (dB) which is a zero mean Gaussian random variable with standard deviation $\delta_{m_h}$. The small scale fading coefficient will be:
\begin{equation}
\beta_{s,{m_h}} = \sqrt{\frac{K_{s,{m_h}}}{1+K_{s,{m_h}}}}+\sqrt{\frac{1}{1+K_{s,{m_h}}}}\Xi_{s,{m_h}},
\end{equation}
where $K_{s_{m_h}}$ is the Rician factor and $\Xi_{s,{m_h}} \sim \mathcal{N}(0,1)$. We can then simplify the channel gain:
\begin{equation}
g_{s,{m_h}}(n) = \Big( \frac{d_0}{d_{s,{m_h}}(n)} \Big)^{\frac{\zeta_{s,{m_h}}}{2}} 10^{-\frac{\omega_{s, {m_h}}+\psi_{s,{m_h}}}{20}}\Bigg( \sqrt{\frac{K_{s,{m_h}}}{1+K_{s,{m_h}}}}+\sqrt{\frac{1}{1+K_{s,{m_h}}}}\Xi_{s,{m_h}} \Bigg)
\end{equation}
The interference at time slot $n$ in this link from non-associated ABSs and CBSs will be $
\Omega_{s,m_h} = \sum\limits_{\forall u \in \mathcal{U} } p_{u} g_{u, m_h} + \sum\limits_{\forall c \in \mathcal{C} } p_{c} g_{c, m_h} $. Here $p_{u}$ is the transmit power of ABS $u$ and $p_{c}$ is the transmit power of CBS $c$. Thus,  the SINR between this link can be given as:
\begin{equation}
\gamma_{s,m_h}(n) = \frac{p_{s,m_h}(n)  g_{s,m_h}(n)}{\Omega_{s,m_h}+\sigma^{2}},\quad  \forall m_h \in \mathcal{M}_h, \forall n \in \mathcal{N},
\end{equation}
where $p_{s,m_h}$ is the transmit of satellite $s$ to HUE $m_h$ over the $z$th  RB at time slot $n$. The achievable data rate between satellite $s$ and HUE $m_h$ in each time slot $n$ can be calculated by Shannon capacity:
\begin{equation}
r_{s,m_h}(n) = B_s\log_2 \left(1 + \gamma_{s,m_h}(n) \right),
\end{equation}
where $B_{m_h}$ is the bandwidth allocated to the channel from satellite $s$ to HUE $m_h$ at time slot $n$.\\

\subsection{CBS-HUEs Data Link Analysis}
Each CBS provides backhaul services to the HUEs near the coastline under their coverage region. Although empirical path loss models can accurately forecast average signal intensity in the marine environment, they are unable to account for the local oscillations caused by the destructive summing of sparse multipath signals. Ray trajectory-based path loss models mathematically detect the trajectories of the most dominating rays arriving at the receiver to solve this problem. As a result, the phase shift of each ray is described and taken into account in the path loss computation, resulting in a more accurate representation of the received signal strength's local peaks and nulls \cite{maritimechannel}. Therefore, path loss between a CBS $c$ and a HUE $m_h$ link can be modeled as curved-earth two-ray (CE2R) which take into account the earth curvature \cite{maritimechannel}:
\begin{equation*}
\xi_{c, {m_h}} = -10 \log_{10} \left\{ \left( \frac {\lambda} {4 \pi d_{c,{m_h}} } \right)^{2} \left[ 2 \sin \left( \frac {2 \pi h_{c} h_{{m_h}}} {\lambda d_{c,{m_h}}} \right) \right]^{2} \right\},
\end{equation*}
where $\xi_{c, {m_h}}$ is the propagation loss in dB, $\lambda$ indicate wavelength of signal, $h_c$ and $h_{m_h}$ is the height of CBS $c$ and HUE $m_h$, respectively. Additionally, $d_{c,{m_h}}(n)$ is the 3D Euclidean distance between CBS $c$ and HUE $m_h$ at each time slot $n$ as:
\begin{equation}
d_{c,{m_h}}(n) = \|\left( \boldsymbol{d}_{c}(n)-\boldsymbol{d}_{{m_h}}(n) \right) \|^{2}, \forall c \in \mathcal{C}, \forall m_h \in \mathcal{M}_h, \forall n \in \mathcal{N},
\end{equation}
The channel gain between this link can be given as:
\begin{equation}
g_{c,{m_h}}(n) = \beta_{c,{m_h}} 10^{-\xi_{c,{m_h}}(n) / 10}, \quad \forall c \in \mathcal{C}, u \in \mathcal{U}, n \in \mathcal{N}.
\end{equation}
The interference at time slot $n$ in this link from ABSs and non-associated CBSs will be $\Omega_{c,m_h} = \sum\limits_{\forall u \in \mathcal{U} } p_u g_{u, m_h} + \sum\limits_{\forall c' \ne c }\sum\limits_{\forall m_h' \ne m_h } p_{c'} g_{c', m_h'}$. Here $c'$ is non-associated CBSs, $p_{c'}$ is the transmit power of non-associate CBSs, and $p_u$ is the transmit power of ABSs. Thus, the SINR between this link will be:
\begin{equation}
\gamma_{c,{m_h}}(n) = \frac{p_{c,{m_h}}(n)  g_{c,{m_h}}(n)}{\Omega_{c,m_h} + \sigma^{2}}, \quad \forall c \in \mathcal{C}, m_h \in \mathcal{M}_h, n \in \mathcal{N},
\end{equation}
where $p_{c,{m_h}}(n)$ is the transmit of CBS's $c$ to HUE over $y$th RB at time slot $n$. The achievable data rate between CBS $c$ and high-end UE $m_h$ in each time slot $n$ can be calculated by Shannon capacity:
\begin{equation}
r_{c,{m_h}}(n) = B_c\log_2 \left(  1 + \gamma_{c,{m_h}}(n)   \right),
\end{equation}
where $B_c$ is the bandwidth allocated to the channel from CBS $c$ to HUE $m_h$ at time slot $n$. \\

\section{Towards an Energy-Efficient Heterogeneous SAS-NTN Maritime Networks}
\label{form}
Our main objective is to provide a decentralized approach that enables the network operator to manage each marine UE and to find its optimal resource allocation based on both its position and user type. Therefore, we seek to maximize the network energy efficiency $\eta$ by factoring in the sum rate $R_t$ and total power $P_t$. Moreover, we need to find the optimal 3D coordinates of the ABSs $\boldsymbol{d}_u$. To realize this, we optimize the position of the ABSs jointly with the marine UEs association $\boldsymbol{a}$ and transmit power control $\boldsymbol{p}$. We formulate the resource allocation and ABSs deployment problem of maximizing the system energy efficiency (Bit/Joule) for the SAS-NTN networks. To formulate this problem, we next define a series of constraints as follows:

Each ABS must return to its initial position at the end of the flight time. This constraint ensures downlink connectivity to LUEs in the marine environment with the pre-defined route and stationary points, so each ABS must travel within the specified area \cite{Sea_routes}:
\begin{equation} \label{C1_UAV}
\boldsymbol{d}_u(1) = \boldsymbol{d}_u(N), \quad \forall u \in \mathcal{U}.
\end{equation}
Then we have the following constraint, which ensures that the distance covered by the ABS between two consecutive time slots corresponds to the distance that can be calculated by the speed and time limits. The ABS's mobility is restricted by its maximum propulsion speed, $v_{\mathrm{max}}$. Furthermore, ABS requires a minimum stall speed $v_{\mathrm{min}}$ in some severe conditions to retain mobility.
\begin{equation}  \label{C2_UAV}
\left\lvert \boldsymbol{d}_{u}[n+1]- \boldsymbol{d}_{u}[n]\right\rvert \leq\left({v_{\max}L_u}\right), \quad \forall u \in \mathcal{U}, \forall n \in \mathcal{N}.
\end{equation}
To ensure the kinematic energy budget for each ABS, the threshold must be met at each time slot $n$ of the flight:
\begin{equation}  \label{C3_UAV}
E_u^{\textrm{flight}}(n) \geq E_{\textrm{th}}(n),  \quad \forall u \in \mathcal{U}, \forall n \in \mathcal{N}.
\end{equation}
The ABS's flight power consumption should be:
\begin{equation} \label{C4_UAV_speed}
P^{\mathrm{flight}}(n) \geq \frac{E_u^{\textrm{flight}}(n)}{L_u},   \quad \forall u \in \mathcal{U}, \forall n \in \mathcal{N}.
\end{equation}
The flight speed of each ABS should be within the range at which the LUEs downlink criterion must be met:
\begin{equation} \label{C5_UAV_speed}
v_{\mathrm{min} }(n) \leq v_{u}(n) \leq v_{\max }(n), \quad \quad \forall u \in \mathcal{U}, \forall n \in \mathcal{N},
\end{equation}
where $v_{\min}(n)$ and $v_{\max}(n)$ denote the minimum and maximum speed of each ABS at the time slot $n$ respectively. The ABS speed limit can be adjusted according to the LUEs requirements \cite{faber2012regulated}. The boundary conditions for each ABS altitude to ensure LoS connections for LUEs have also been established:
\begin{equation} \label{C6_UAV_height}
h_{\min }(n) \leq h_{u}(n) \leq h_{\max }(n), \quad \quad \forall u \in \mathcal{U}, \forall n \in \mathcal{N},
\end{equation}
where $h_{\min}$ ensures a LoS link between the ABS and LUEs, and $h_{\max}$ is an upper bound defined by air traffic control \cite{capacity}. It is considered that each ABS can utilize the satellite (space-to-air) or any CBS (coastline-to-air) for backhaul connectivity. The aggregated achievable rate of all ABSs-to-LUEs links should remain within the channel capacity of satellite-to-ABS and CBS-to-ABS links. These constraints guarantee the capacity of the backhaul as follows:
\begin{equation} \label{C7_UAV_SATELLITE_DATA_CAPACITY}
\begin{aligned}
& \sum_{u=1}^{U} \sum_{m_{l}=1}^{M_{l}}  r_{u, m_{l}}(n)  \leq \sum_{u=1}^{U} r_{u, s}(n), \quad \forall n \in \mathcal{N},
\end{aligned} 
\end{equation}		
\begin{equation} \label{C8_UAV_CBS_DATA_CAPACITY}
\begin{aligned}		
&\sum_{u=1}^{U} \sum_{m_{l}=1}^{M_{l}} r_{u, m_{l}}(n)  \leq \sum_{c=1}^{C} \sum_{u=1}^{U}  r_{c, u}(n), \quad \forall n \in \mathcal{N}.
\end{aligned}
\end{equation}
Each ABS $u$ need to satisfy the demand of each associated  LUEs data rate which can be defined as:
\begin{equation} \label{C9_UAV_DATA_DEMAND_SATISFACTION}
\begin{aligned}
& \sum_{n \in \mathcal{N}} a_{u, m_{l}} r_{u, m_{l}}(n)  \geq r_{\mathrm{th}}, \quad \forall u \in \mathcal{U}, \forall m_l \in \mathcal{M}_l,
\end{aligned}
\end{equation}
where $r_{\mathrm{th}}$ is the minimum data rate requirement of each LUE. The downlink transmit power of each ABSs $u$ for associated LUEs should be remain within the power budget limits:
\begin{equation}  \label{C10_UAV_TRANSMIT POWER}
\begin{aligned}  
& 0 \leq a_{u,m_l} p_{u,m_l}(n) \leq p_{\max}, & \forall u \in \mathcal{U}, m_l \in \mathcal{M}_l, n \in \mathcal{N},
\end{aligned}
\end{equation}
To ensure a safe distance between the $U$ ABS, we define a secure distance that can avoid an overlap in their coverage region. This threshold distance can be defined for all ABSs $\forall i,j \in \mathcal{U}$:
\begin{equation} \label{C11_UAV_distance_secure} 
\begin{aligned}  
&  \| \boldsymbol{d}_{i}(n)- \boldsymbol{d}_{j}(n) \|^2 \geq d_{\mathrm{th}}, \quad \forall {i, j \in \mathcal{U}, i \neq j}.
\end{aligned}  
\end{equation}
Each ABS $u$ can assign each resource block $k$ at each time slot $n$ to a maximum of one LUE that can be given as:
\begin{equation} \label{C12_UAV_LOW_UE_ASSOCIATION}
\begin{aligned}
& \sum_{u=1}^{U} \sum_{k=1}^{K} \sum_{m_{l}=1}^{M_{l}} a_{u, k, m_{l}}(n) \leq 1,\forall n \in \mathcal{N}, \\
& a_{u, k, m_{l}}(n) \in\{0,1\}, \quad \forall u \in \mathcal{U}, \forall k \in \mathcal{K},\forall m_l \in \mathcal{M}_l.
\end{aligned}
\end{equation}
In addition, each ABS $u$ can be associated with atmost one backhaul service node, depending on its position in the sea, which can be defined as:
\begin{equation} \label{C13_UAV_BACKHAUL_ASSOCIATION}
\begin{aligned}
& \sum_{u=1}^{U} a_{s, u}(n) \leq 1, \quad  a_{s, u}(n) \in\{0,1\}, \quad \forall n \in \mathcal{N},\\
& \sum_{c=1}^{C} \sum_{u=1}^{U} a_{c, u}(n) \leq 1, \quad a_{c, u}(n) \in\{0,1\}, \quad \forall n \in \mathcal{N},
\end{aligned}
\end{equation}
The satellite $s$ can assign each resource block $z$ at each time slot $n$ to a maximum of one HUE or ABS for backhaul that can be given as respectively:
\begin{equation} \label{C14_SATELLITE_HIGH_UE_ASSOCIATION}
\begin{aligned}
& \sum_{z=1}^{Z} \sum_{m_h=1}^{{M}_h} a_{s, z, m_{h}}(n) \leq 1, \quad \forall n \in \mathcal{N}, \\
& a_{s, z, m_{h}}(n) \in \{0,1\}, \quad \forall k \in \mathcal{K},\forall m_h \in \mathcal{M}_h,
\end{aligned}
\end{equation}
\begin{equation} \label{C15_SATELLITE_UAV_BACKHAUL_ASSOCIATION}
\begin{aligned}
& \sum_{z=1}^{Z} \sum_{u=1}^{U} a_{s, z, u}(n) \leq 1,\quad \forall n \in \mathcal{N}, \\
& a_{s, z, u}(n) \in \{0,1\}, \quad \forall k \in \mathcal{K},\forall u \in \mathcal{U}.
\end{aligned}
\end{equation}
Similarly, each CBS $c$ can assign each resource block $y$ at each time slot $n$ to a maximum of one HUE or ABS for backhaul that can be given as respectively:
\begin{equation} \label{C16_CBS_HIGH_UE_ASSOCIATION}
\begin{aligned}
& \sum_{c=1}^{C}\sum_{y=1}^{Y}\sum_{m_h=1}^{{M}_h} a_{c, y, m_{h}}(n) \leq 1, \quad \forall n \in \mathcal{N}, \\
& a_{c, y, m_{h}}(n) \in \{0,1\}, \quad \forall c \in \mathcal{C}, y \in \mathcal{Y},\forall m_h \in \mathcal{M}_h,
\end{aligned}
\end{equation}
\begin{equation} \label{C17_CBS_UAV_BACKHAUL_ASSOCIATION}
\begin{aligned}
& \sum_{c=1}^{C} \sum_{y=1}^{Y} \sum_{u=1}^{U} a_{c, y, u}(n) \leq 1,\quad \forall n \in \mathcal{N}, \\
& a_{c, y, u}(n) \in \{0,1\}, \quad \forall c \in \mathcal{C}, \forall y \in \mathcal{Y},\forall u \in \mathcal{U}.
\end{aligned}
\end{equation}
The large transmission distances between the satellite and ABSs are assumed to be constant at each time slot $n$ due to the short interval. The satellite and CBSs must meet the downlink demand of associated HUEs i.e.:
\begin{equation}  \label{C18_SATELLITE_H_UE_DATARATE}
\begin{aligned} 
& \sum_{n \in \mathcal{N} } a_{s,m_h} r_{s, m_{h}}(n) \geq r_{\mathrm{th}}, \quad \forall m_h \in \mathcal{M}_h,
\end{aligned}
\end{equation}
\begin{equation}  \label{C19_CBS_H_UE_DATARATE}
\begin{aligned}
& \sum_{n \in \mathcal{N} } a_{c,m_h} r_{c, m_{h}}(n) \geq r_{\mathrm{th}}, \quad \forall c \in \mathcal{C}, \forall m_h \in \mathcal{M}_h,
\end{aligned}
\end{equation}
where $r_{\mathrm{th}}$ is each HUE datarate requirement threshold, respectively. Similarly, the downlink transmit power of satellite $s$ for associated devices in each $z$ RB should remain within the power budget limits:
\begin{equation}  \label{C20_SATELLITE_UAV_TRANSMIT_POWER}
\begin{aligned}
& 0 \leq a_{s,u} p_{s,u}(n) \leq p_{\max }, & \forall n \in \mathcal{N}, \forall u \in \mathcal{U},\\
\end{aligned}
\end{equation}
\begin{equation}   \label{C21_SATELLITE_H_UE_TRANSMIT_POWER}
\begin{aligned}
& 0 \leq a_{s,m_h} p_{s,m_h}(n) \leq p_{\max}, & \forall m_h \in \mathcal{M}_h ,\forall n \in \mathcal{N}.
\end{aligned}
\end{equation}
The downlink transmit power of each CBS $c$ for associated devices in each $y$ RB should remain within the power budget limits:
\begin{equation}   \label{C22_CBS_UAV_TRANSMIT_POWER}
\begin{aligned}
& 0 \leq a_{c,u} p_{c,u}(n) \leq p_{\max }, & \forall c \in \mathcal{C}, \forall u \in \mathcal{U}, \forall n \in \mathcal{N},
\end{aligned}
\end{equation}
\begin{equation}   \label{C23_CBS_H_UE_TRANSMIT_POWER}
\begin{aligned}
& 0 \leq a_{c,m_h} p_{c,m_h}(n) \leq p_{\max}, & \forall c \in \mathcal{C}, \forall m_h \in \mathcal{M}_h, \forall n \in \mathcal{N}.
\end{aligned}
\end{equation}   
\subsection{Problem Formulation}
Given the network specifics described above, our objective is to establish an efficient allocation of resources and a maritime UE association scheme that will maximize the EE of the network while meeting the request for user data services within a limited period. We can define the total network EE (Bit/Joule) as follows:
\begin{equation}
\label{eta_defination}
\eta_{\textrm{EE}}(n) = \frac{{R_t}(n)}{{P_t}^+(n)},
\end{equation}
where $R_t$ indicates the total data rate and $P_t^+$ indicates the non-negative power needed to transmit this data and operate the network nodes at time slot $n$, For the sake of understanding, we can define a separate EE for each network node. The EE of $U$ ABSs at time slot $n$ can be defined as follows:
\begin{equation}   \label{UAV_objective_first_simple}
\eta_{u}(n) = \frac{{R_u}(n)}{{P_u}^+(n)} =  \sum_{\mathcal{U}} \sum_{\mathcal{M}_l}     \left( \frac{r_{u,m_l}(n) }{p_{u,m_l}(n) + p^\mathrm{flight}} \right), \quad \forall n \in \mathcal{N}.
\end{equation}
The EE of satellite $s$ at time slot $n$ can be defined as:
\begin{multline}
\eta_{s}(n) = \frac{{R_s}(n)}{{P_s}^+(n)} = \\ \sum_{\mathcal{U}}\sum_{\mathcal{M}_h} \left(  \frac{r_{s,u}(n)+r_{s,m_h}(n)}{p_{s, u}(n)+p_{s, m_h}(n)+ p^{\textrm{circuit}}_s} \right) \quad \forall n \in \mathcal{N}.
\end{multline}
Similarly, the EE of $C$ CBS at time slot $n$ can be stated as follows:
\begin{multline}
\eta_{c}(n) = \frac{{R_c}(n)}{{P_c}^+(n)} = \\ \sum_{\mathcal{C}} \sum_{\mathcal{U}} \sum_{\mathcal{C}} \left( \frac{r_{c,u}(n)+r_{c,m_h}(n) }{p_{c, u}(n)+p_{c, m_h}(n)+p^{\textrm{circuit}}_c} \right), \quad \forall n \in \mathcal{N}.
\end{multline}
Thus, the total network EE can now be define as:
\begin{equation}
\eta_{\textrm{EE}}(n) = \eta_{u}(n)+\eta_{s}(n)+\eta_{c}(n),\quad \forall n \in \mathcal{N}.
\end{equation}
According to the above analysis, the optimization problem of both HUEs and LUEs association, resource allocation, and ABSs deployment for maximizing the SAS network EE can be formulated as follows:
\begin{equation}    \label{optimize}
\begin{aligned} 
&     &         \underset{\boldsymbol{a},\boldsymbol{p},\boldsymbol{d}_{u}}{\text{max}}
&     &   &     \eta_{\mathrm{EE}}, \\
&     &         \text{s.t.}
&     & &       (\ref{C1_UAV})-(\ref{C23_CBS_H_UE_TRANSMIT_POWER}),
\end{aligned}
\end{equation}
where $\mathrm{{\eta}_{EE}}$ is given in (\ref{eta_defination}). The objective function in (\ref{optimize}) is a function of users' association $\boldsymbol{a}$, transmission power $\boldsymbol{p}$, and the ABS $3$D deployment $\boldsymbol{d_n}$. In the given problem, the UEs association constraints in (\ref{C12_UAV_LOW_UE_ASSOCIATION}), (\ref{C14_SATELLITE_HIGH_UE_ASSOCIATION}), and (\ref{C16_CBS_HIGH_UE_ASSOCIATION}) are integer (binary) constraints. Similarly, the ABS selection constraints in (\ref{C13_UAV_BACKHAUL_ASSOCIATION}), (\ref{C15_SATELLITE_UAV_BACKHAUL_ASSOCIATION}), and (\ref{C17_CBS_UAV_BACKHAUL_ASSOCIATION}), are also integer constraints, and the objective function in (\ref{optimize}) is in fractional form, which makes this problem a mixed integer non-convex fractional optimization problem. Moreover, the problem is combinatorial due to the association (binary) constraints in (\ref{C12_UAV_LOW_UE_ASSOCIATION}), (\ref{C14_SATELLITE_HIGH_UE_ASSOCIATION}), and (\ref{C16_CBS_HIGH_UE_ASSOCIATION}). In fact, this problem is a non-deterministic polynomial-time hard (NP-hard) problem.\\

\section{Proposed Solution}
\label{decom}
In this section, we will present our proposed algorithm based on the BD, DA, ADMM, and Gurobi optimizer \cite{gurobi_otimizer}. We developed our algorithm architecture based on the BD structure. Then we solve the master problem by using the Gurobi optimization solver. In the sub-problem, the DA is used to handle fractional programming. We use ADMM to provide a distributed solution in the inner loop of the DA. Details are given in the following subsections.
	
The main challenge of solving the problem (\ref{optimize}) is the non-concavity caused by the fractional form of the objective function and non-convexity due to maritime UEs association (binary) variables constraints given in (\ref{C12_UAV_LOW_UE_ASSOCIATION}), (\ref{C14_SATELLITE_HIGH_UE_ASSOCIATION}), and (\ref{C16_CBS_HIGH_UE_ASSOCIATION}), and ABSs backhaul selection variable given in (\ref{C13_UAV_BACKHAUL_ASSOCIATION}), (\ref{C15_SATELLITE_UAV_BACKHAUL_ASSOCIATION}), and (\ref{C17_CBS_UAV_BACKHAUL_ASSOCIATION}). In order to obtain the solution to this problem, we first decompose (\ref{optimize}) into three subproblems by taking advantage of its block separability.%, or trivially parallelizable structure \cite{conejo2006decomposition} such that ABSs EE (\ref{optimize_UAV_1}), satellite EE (\ref{optimize_SATELLITE_1}), and CBSs EE (\ref{optimize_CBS_1}) individually. 
Thus, the first subproblem is established for the ABS EE at each time slot $n$, as follows:
\begin{equation}  \label{optimize_UAV_1}
\begin{aligned} 
&       &       \underset{\boldsymbol{a}_u,\boldsymbol{p}_u,\boldsymbol{d}_{u}}{\text{max}}
&       &       &       {\eta_{u}(n)}, \\
&       &       \text{s.t.}
&       &       &       (\ref{C1_UAV})-(\ref{C13_UAV_BACKHAUL_ASSOCIATION}),
\end{aligned}
\end{equation}
The second subproblem is established for the satellite EE at each time slot $n$, as follows:
\begin{equation}   \label{optimize_SATELLITE_1}
\begin{aligned} 
&       &       \underset{\boldsymbol{a}_s,\boldsymbol{p}_s}{\text{max}}
&       &       &   \eta_{s}(n), \\
&       &       \text{s.t.}
&       &       &     (\ref{C14_SATELLITE_HIGH_UE_ASSOCIATION}), (\ref{C15_SATELLITE_UAV_BACKHAUL_ASSOCIATION}), (\ref{C18_SATELLITE_H_UE_DATARATE}),(\ref{C20_SATELLITE_UAV_TRANSMIT_POWER}), (\ref{C21_SATELLITE_H_UE_TRANSMIT_POWER}).
\end{aligned}
\end{equation}
The third subproblem is established for the CBSs EE at each time slot $n$, as follows:
\begin{equation}   \label{optimize_CBS_1}
\begin{aligned} 
&           &       \underset{\boldsymbol{a}_c,\boldsymbol{p}_c}{\text{max}}
&           &       &       \eta_{c}(n), \\
&           &       \text{s.t.}
&           &       &       (\ref{C16_CBS_HIGH_UE_ASSOCIATION}), (\ref{C17_CBS_UAV_BACKHAUL_ASSOCIATION}), (\ref{C19_CBS_H_UE_DATARATE}), (\ref{C22_CBS_UAV_TRANSMIT_POWER}), (\ref{C23_CBS_H_UE_TRANSMIT_POWER}).
\end{aligned}
\end{equation}
We next tackle each problem individually.\\

\subsection{Aerial Base Stations Energy Efficiency (ABSs-EE)}
This part introduces an optimization scheme of LUEs' association, transmit power control, and ABSs' deployment for (\ref{optimize_UAV_1}). This optimization algorithm describes maximizing the ABSs' energy efficiency in the SAS-NTN networks based on BD, DA, ADMM, and optimization solver.

\subsubsection{Bender Decomposition for ABS EE}
The BD algorithm is a solution approach for tackling constraints in optimization problems based on the idea of partition and delayed constraint generation \cite{conejo2006decomposition}. Firstly, a mathematical problem formulation is proposed \cite{conejo2006decomposition} as MINLP, then decompose the problem in two parts:
\begin{itemize}
\item A master problem, which deals with binary constraints by branch and bound (B\&B) technique, finds values for a subset of the original variables and associated constraints.
\item One or more subproblems are used to find the solution for the remaining original variables by any linear programming (LP) method while keeping the master problem variables constant.
\end{itemize}
Both problems are solved iteratively until convergence. In the master problem, there are some added constraints called the Benders Cut to cut the solution region. When the upper and lower bounds meet or the difference between them is lower than a certain threshold, the optimal solution will be given.\\
\indent \textbf{Initialization:} We first assume that the master problem has a trivial solution and can be solved by generating the initialization in the given problem. Then, we need to assign the loop counter, i.e., ${i_u} = 1$. In our problem, we have an association variable $a_u$ in binary form, and, thus, the upper and lower bounds will be $a_{\textrm{UB}}=1$ and $a_{\textrm{LB}}=0$ respectively. Moreover, we implement a function $\chi_u$ as an auxiliary variable, representing the objective function of the subproblems within the objective function of the master problem. We can set the initial value for a function $\chi_u^{\psi}$ as $\chi_u^{\textrm{down}}$, to avoid an unbounded solution in the first iteration when there is no cut in the master problem. It can be initiated with a negative value, i.e., $-10^6$.\\
\indent \textbf{Subproblems:} The idea behind the construction of subproblems is to fix the value of association variables $\boldsymbol{a}_u$ to avoid them. Therefore, we can express the subproblem as given in (\ref{SP_UAV}). We can represent the dual variable for the fronthaul constraints in each ABS $u$ that fixed association variables values, i.e., $\kappa_{u, m_l}^{i_u}$ from ABS $u$ to LUE $m_l$ and backhaul constraints $\kappa_{s, u}^{i_u}$ and $\kappa_{c, u}^{i_u}$ form the satellite $s$ and CBS $c$ to ABS $u$ respectively. Hence, the subproblem can be obtained with only transmit power and ABSs' deployment continuous variables, and it can be represented as:
\begin{subequations} \label{SP_UAV}
\begin{align}
\underset{\boldsymbol{p}_u, \boldsymbol{d}_u}{\text{max}} \quad &  \tilde{\eta}_u, \\
\text {s.t.} \quad & \tilde{\eta}_u = \frac{R_{u}\left(\tilde{\boldsymbol{a}}_u, \boldsymbol{p}_u, \boldsymbol{d}_{u}\right)}{P_{u}^{+}},\label{SP_UAV_C1}\\
							 & \boldsymbol{a}_{u, m_l} = \boldsymbol{a}_{u, m_l}^{i_u}: \kappa_{u, m_l}^{{i_u}}, \quad u \in \mathcal{U}, \forall m_l \in \mathcal{M}_l, \label{SP_UAV_C2}  \\
							 & \boldsymbol{a}_{s,u} = \boldsymbol{a}_{s,u}^{{i_u}} : \kappa_{s,u}^{{i_u}}, \quad  \forall u \in \mathcal{U}, \label{SP_UAV_C3}\\
							 & \boldsymbol{a}_{c,u} = \boldsymbol{a}_{c,u}^{{i_u}} : \kappa_{c,u}^{{i_u}}, \quad  c \in \mathcal{C},\forall u \in \mathcal{U},  \label{SP_UAV_C4}\\
							 & (\ref{C1_UAV}) - (\ref{C11_UAV_distance_secure}) \label{SP_UAV_C5},
\end{align}
\end{subequations}
where $\boldsymbol{\tilde{a}_u}$ the fixed value of each association vector from the initial master problem solution, and this fixing value constraint is stated in (\ref{SP_UAV_C2}), (\ref{SP_UAV_C3}), and (\ref{SP_UAV_C4}). After solving this subproblem, we will get the sub-optimal transmit power $\boldsymbol{p}_u^*$ and the deployment vector $\boldsymbol{d}_n^*$\footnote{Hereinafter, the ABSs' deployment vector can be alternatively used with these notations, i.e., $\boldsymbol{d}_u = \{x_u, y_u\}$.} for each ABS $u$. This obtained subproblem will be solved by utilizing DA in Section \ref{Dinkelbach}. \\
\indent \textbf{Convergence Analysis and Bounds:} This process is used to derive upper and lower bounds that are used as the stopping criterion for the algorithm and as a condition for the convergence. In this step, we obtain the upper and lower bound difference. The objective function at iteration ${i_u}$ provides the upper bound, which is stated as:
\begin{equation}
\label{upper_bound_ABS}
\eta_{\mathrm{UB}}^{{i_u}}=\frac{\tilde{R}_{u}\left(\boldsymbol{a}_u^{{i_u}}, \boldsymbol{p}_u^{{i_u}}, \boldsymbol{d}_{u}^{{i_u}}\right)}{\tilde{P}_{u}\left(\boldsymbol{a}_u^{{i_u}}, \boldsymbol{p}_u^{{i_u}}, \boldsymbol{d}_{u}^{{i_u}}\right)^+},
\end{equation}
where $\tilde{R}_{u}$ and $\tilde{P}_{u}$ are intermediate values of both parameters at iteration $i_u$ which depend on the sum rate from all associated LUEs and transmit power consumption respectively. The lower bound can be given as:
\begin{equation}
\label{lower_bound_ABS}
\eta_{\mathrm{LB}}^{{i_u}} = \chi_u^{{i_u}}.	
\end{equation}
Therefore, the stopping criterion can be stated as:
\begin{equation}
\left\{\begin{array}{ll}
\eta_{\mathrm{UB}}^{{{i_u}}}-\eta_{\mathrm{LB}}^{{{i_u}}} \leq \epsilon, & \textbf {stop}, \\
\text {otherwise}, & \textbf {continue},
\end{array}\right.
\end{equation}
where $\epsilon$ is a pre-defined tolerance parameter. Thus, after convergence the sub-optimal values of $\boldsymbol{a}_u^*$, $\boldsymbol{p}_u^*$ and $\boldsymbol{d}_u^*$ can be obtained.\\
\indent \textbf{Master Problem:} This problem deals only with association variables while all other variables remain fixed. The loop counter can be update as ${{i_u}} = {{i_u}}+1$, and after that the solvable problem become as follows:
\begin{subequations} \label{MP_UAV}
\begin{align}
\underset{\boldsymbol{a}_u, {\chi}_u}{\text{max}} \quad 	& {\chi}_u, \\
\text {s.t.} 	\quad								  	    & (\ref{C12_UAV_LOW_UE_ASSOCIATION}), (\ref{C13_UAV_BACKHAUL_ASSOCIATION})\\
														    & (\ref{bender_cut_MP})\label{MP_CUT},\\
															& \chi_u  \geq \chi^{\text {down }},
\end{align}
\end{subequations}
where inequality constraint in (\ref{MP_CUT}) represents the Bender cut in the master problem. At every iteration, the new Benders cut will generate and append to the master problem. Additionally, the previous iteration's Bender cuts remain the same in the master problem. The master problem becomes the mixed integer programming problem which only decides the associations and this can be solved with an optimization solver to reduce the complexity. At each iteration, we obtain the optimal values of association $\boldsymbol{a}_u^*$ and auxiliary variable $\chi_u$.

After each iteration of the master problem, we solve the subproblem again using the obtained local optimal values. Therefore, when the optimal criterion of upper and lower is met, the iteration process will stop. These details of the Benders technique are presented in Algorithm \ref{Algorithm_1}. After getting the optimal user association $\boldsymbol{a}^*_u$, the subproblem is still non-convex due to its objective function. Note that the objective function is non-convex in (\ref{SP_UAV_C1}). Therefore, we apply the Taylor approximation to the numerator term in (\ref{SP_UAV_C1}) to linearize the objective function as given in equation (\ref{taylorexpansion1}). 
\begin{lemma}
Since the first-order Taylor approximation is the global lowest bound of a convex function and the global upper bound of a concave function \cite{taylorseries}. 
\end{lemma}
\indent \textit{Proof:} See Appendix A.
\begin{figure*}[b] 
\noindent\rule{\textwidth}{0.4pt}\\
\small
\begin{equation} \label{taylorexpansion1}
\begin{aligned}
R_{u,m_l}^{lb} = 
\sum_{u=1}^U \sum_{m_{l}=1}^{M_l} \tilde{a}_{u,m_l}B_{u,m_l} \left[ \log_2 \bigg\{ 1 + \frac{p_{u,m_l}g_0}{(\Omega_{u,m_l} + \sigma^{2}) \Big( \Big\|  \big( \boldsymbol{d}_{u, \mathrm{local}}(n)-\boldsymbol{d}_{m_{l}}(n) \big) \Big\|^{2} \Big) } \bigg \}- 
\frac{ p_{u,m_l}g_0 \Big\{ \Big\| \big( \boldsymbol{d}_{u}(n)-\boldsymbol{d}_{m_{l}}(n) \big) \Big\|^{2} - \Big\| \big( \boldsymbol{d}_{u, \mathrm{local}  }(n)-\boldsymbol{d}_{m_{l}}(n) \big) \Big\|^{2} \Big\} \log_2 e}{ \Big\{ \Big\| \big( \boldsymbol{d}_{u, \mathrm{local}}(n)-\boldsymbol{d}_{m_{l}}(n) \big) \Big\|^{2} \Big\} \Big\{ p_{u,m_l}g_0 +  (\Omega_{u,m_l} + \sigma^{2}) \Big( \Big\| \big( \boldsymbol{d}_{u, \mathrm{local}}(n)-\boldsymbol{d}_{m_{l}}(n) \big) \Big\|^{2} \Big) \Big\} }  \right]  
\end{aligned}
\end{equation}
\small
\begin{equation} \label{bender_cut_MP}
\begin{aligned}
{\chi}_u \leq \eta_{\mathrm{UB}}^{{i_u}}+ \sum_{u=1}^{U}\sum_{m_{l}=1}^{M_{l}} \kappa_{u,m_{l}}^{{i_u}}\left(\boldsymbol{a}_{u,m_{l}}-\boldsymbol{a}_{u,m_{l}}^{{i_u}}\right)  +\sum_{u=1}^{U} \kappa_{s,u}^{{i_u}}\left(\boldsymbol{a}_{s,u}-\boldsymbol{a}_{s,u}^{{i_u}}\right)  +\sum_{c=1}^{C}\sum_{u=1}^{U} \kappa_{c,u}^{{i_u}}\left(\boldsymbol{a}_{c,u}-\boldsymbol{a}_{c,u}^{{i_u}}\right),
\end{aligned}
\end{equation}
\small
\begin{equation} \label{taylorexpansion2}
\begin{aligned}
f(\boldsymbol{d}_u)^{lb} = \big\| \boldsymbol{d}_{u1, \mathrm{local}} - \boldsymbol{d}_{u2, \mathrm{local}} \big\|^2 + 2 \big( \boldsymbol{d}_{u1, \mathrm{local}} - \boldsymbol{d}_{u2, \mathrm{local}}  \big) \big( \boldsymbol{d}_{u1}- \boldsymbol{d}_{u2} \big)^T.
\end{aligned}
\end{equation}
\end{figure*}
\begin{algorithm}[t]
\caption{Outer Loop Bender Decomposition}
\label{Algorithm_1}
\begin{algorithmic}[1]
\State \textbf{Input:} Initialize variables $\tilde{\boldsymbol{a}}$, loop counter ${i_u}$, $\chi_u^\psi = \chi^{\textrm{down}} $
\State \textbf{Output:} optimal solution $\boldsymbol{a}_u^*$, $\boldsymbol{p}_u^*$, and $\boldsymbol{d}_n^*$
\While {$\eta_{\mathrm{UB}}^{{i_u}}-\eta_{\mathrm{LB}}^{{i_u}} \geq \epsilon$} 
\State \textbf{Subproblem}
\State \indent obtain $\boldsymbol{p}_u^*$ and $\boldsymbol{d}_n^*$ using Dinkelbach algorithm
\State \textbf{Bounds calculation}
\State \indent calculate upper and lower bounds $\eta_{\mathrm{UB}}^{{i_u}}$ and $ \eta_{\mathrm{LB}}^{{i_u}}$ 
\indent \hspace{4 mm} by (\ref{upper_bound_ABS}) and (\ref{lower_bound_ABS})
\State \textbf{Master Problem}
\State \indent step 1: Increment in loop counter ${i_u} = {i_u} + 1$
\State \indent step 2: Add the new Bender cut in (\ref{MP_UAV})
\State \indent step 3: Solve the updated master problem in (\ref{MP_UAV})
\State \indent step 4: Acquire the optimal value $\boldsymbol{a}_u^*$ and $\chi_u$
\EndWhile
\end{algorithmic}
\end{algorithm}
\subsubsection{Dinkelbach Algorithm for ABS EE}
\label{Dinkelbach}
We use the DA to address the fractional nature of the objective function. Fortunately, this method will always converge to local optima \cite{dinkelbach1967nonlinear}. The DA is widely adopted in solving the fractional programming \cite{green_fog_fractional}. It can be observed from (\ref{SP_UAV}) that it has a fractional objective function. Therefore, we can employ nonlinear fractional programming to transform the original problem in fractional from into an equivalent subtractive form. Without loss of generality, the system maximum average EE can be given as:
\begin{equation}
\tilde{\eta}_u^{*}=\frac{R_{u,m_l}^{lb}\left(\boldsymbol{\tilde{a}_u^{*}}, \boldsymbol{p}_u^{*}, \boldsymbol{d}_{u}^{*}\right)}{P_{u}^{+}}= \underset{ \boldsymbol{p}_u,\boldsymbol{d}_{u}}{\arg \max} {\frac{R_{u,m_l}^{lb}\left(\boldsymbol{\tilde{{a}}}_u^{*}, \boldsymbol{p}_u, \boldsymbol{d}_{u}\right)}{P_{u}^{+}}},
\end{equation}
then, we introduce a Remark \ref{remark_1} to solve the optimization problem in (\ref{SP_UAV}).
\begin{remark}
\label{remark_1}
When $R_{u,m_l}^{lb}\left(\boldsymbol{\tilde{{a}}}_u, \boldsymbol{p}_u, \boldsymbol{d}_{u}\right) \geq 0$ and $P_{u}^+ > 0 $ is fulfilled, the objective function in (\ref{SP_UAV}) can be rewritten to a parametric subtractive form equivalently if and only if the following condition is satisfied:
\begin{equation}
\begin{aligned}
\max _{\boldsymbol{p}_u, \boldsymbol{d}_{u}} \quad &     R_{u,m_l}^{lb}\left(\boldsymbol{\tilde{a}}_u^{*}, \boldsymbol{p}_u, \boldsymbol{d}_{u}\right)-\eta_u^{*} P_{u}^{+} & \\
\quad &=R_{u,m_l}^{lb}\left(\boldsymbol{\tilde{a}}_u^{*}, \boldsymbol{p}_u^{*}, \boldsymbol{d}_{u}^{*}\right)-\eta_u^{*} P_{u}^{+}.
\end{aligned}
\end{equation}
\end{remark}
This Remark \ref{remark_1} illustrates that there exists an equivalent transformed problem with an objective function in subtractive form, which leads to the same maximum $\eta_u^*$ obtained by directly solving (\ref{SP_UAV}). Our objective function is a strictly monotonic increasing function of $\eta_u$ which can be stated as:
\begin{equation}
F(\boldsymbol{a}_u,\boldsymbol{p}_u, \boldsymbol{d}_u; \eta_u) = R_{u,m_l}^{lb}\left(\boldsymbol{\tilde{a}}_u, \boldsymbol{p}_u, \boldsymbol{d}_{u}\right)-\eta_u P_{u}^{+}.
\end{equation}
Thus, the equivalent optimization problem in subtractive form is reformulated as:
\begin{equation}
\label{modified_SP_1}
\begin{aligned}
\underset{\boldsymbol{p}_u, \boldsymbol{d}_u}{\text{max}} \quad &     F(\boldsymbol{\tilde{a}}_u,\boldsymbol{p}_u, \boldsymbol{d}_u; \eta_u), \\
\text { s.t. } \quad & (\ref{SP_UAV_C1}) - (\ref{SP_UAV_C5}).
\end{aligned}
\end{equation}
The nonlinear fractional objective function is transformed into a subtractive objective function, which is a multi-objective convex optimization problem whereby the variable $\eta_u$ (non-negative) can be regarded as a negative weight of $\boldsymbol{p}_u$.At last, parameter $\eta_u$ updates itself after each iteration and finally obtains the sub-optimality condition, which can be defined as $\eta_u^*$. The details of DA are provided in Algorithm \ref{algorithm_2}. The safe distance constraint is given in (\ref{C11_UAV_distance_secure}) between ABSs is of the quadratic type. Therefore, we provide the following lemma \ref{lemma2} to linearize it.
\begin{lemma} \label{lemma2}
 We can linearize this constraint by approximating it with first-order Taylor expansion, which can also be the lower bound for the distance threshold as given in (\ref{taylorexpansion2}).
\end{lemma}
\indent \textit{Proof:} See Appendix B.
\begin{algorithm}[t]
\caption{Inner Loop Dinkelbach Algorithm}
\label{algorithm_2}
\begin{algorithmic}[1]
\State \textbf{Input:} loop counter $j=0$, energy efficiency $\eta_u=0$, maximum tolerance $\Upsilon$
\State \textbf{Output:} optimal power $\boldsymbol{p}_u^{*}$, and UAVs deployment $\boldsymbol{d}_u^{*}$
\State Maximum energy efficiency $= \eta_u^*$
\While { $  \| F(\boldsymbol{a}_u,\boldsymbol{p}_u, \boldsymbol{d}_u; \eta_u) \geq   \Upsilon \|   $ }
\State Solve the subproblem ($\ref{modified_SP_1}$) by ADMM to find the the \indent \kern-0.5em optimal solution $\boldsymbol{p}_u^{*}$ and $\boldsymbol{d}_u^{*}$ with $\eta_u$ 
\State Calculate $\eta_u = \frac{R_u}{P^+_u}$ with obtained $\boldsymbol{p}_u^{*}$ and $\boldsymbol{d}_u^{*}$
\State Calculate new $F(\boldsymbol{a}_u,\boldsymbol{p}_u, \boldsymbol{d}_u; \eta_u) $ with updated $\eta_u$, $\boldsymbol{p}_u^{*}$ \indent \kern-0.5em and $\boldsymbol{d}_u^{*}$
\State Update loop counter $j = j + 1$
\EndWhile
\end{algorithmic}
\end{algorithm}
\subsubsection{ADMM for ABS EE}
For the subproblem, we use ADMM to solve it in a distributed way. An ADMM is commonly used to decouple the constraint linked with all ABSs. The original problem also costs a great deal of time and resources. By splitting the problem into small problems, time and money can be saved in green communication. 

\begin{figure}[t]  
\centering
\includegraphics[width=\columnwidth]{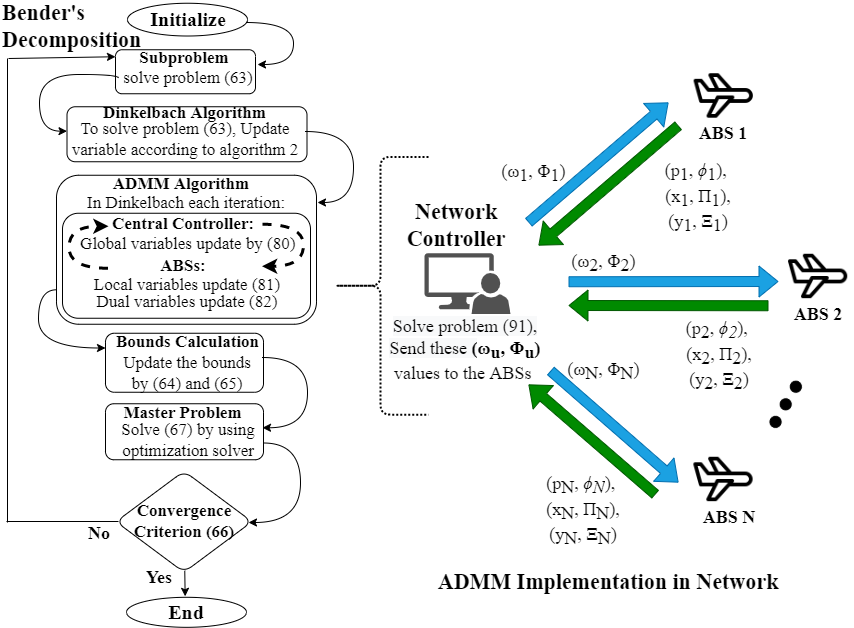}
\caption{Illustration of Bender decomposition, Dinkelbach algorithm, and ADMM working procedure for ABSs.}
\label{fig:whole_flowchart}
\vspace{-0.25in}
\end{figure}

Firstly, we need to turn (\ref{modified_SP_1}) into a solvable problem. In this subproblem, we introduce three auxiliary variables $\omega_u$, $\nu_{u}$ and $o_{u}$ as global copies, which implies that three new equality constraints are applied to the subproblem (\ref{modified_SP_1}), which can be given as:
\begin{subequations}
\begin{align}
\boldsymbol{p}_{u} = \boldsymbol{\omega}_{u}, \quad \forall u \in \mathcal{U}, \label{equality_constraint_pow_1}\\
x_{u} = \nu_{u}, \quad \forall u \in \mathcal{U}, \label{equality_constraint_x_1}\\
y_{u} = o_{u}, \quad \forall u \in \mathcal{U} \label{equality_constraint_y_1},
%f(\boldsymbol{d}_u)^{lb} = f(\boldsymbol{\nu}_{u})^{lb}, \quad \forall u \in \mathcal{U}.
\end{align}
\end{subequations}
where $\omega_{u}$ is the global copy of transmit power variables. Similarly, $\nu_{u}$ and $o_{u}$ are the global copies of $x$ and $y$ coordinates' decision variables for each ABS deployment, respectively. Therefore, the ABS's deployment vector $\boldsymbol{d}_u$ in a global problem can be represented by $\boldsymbol{\Theta}_u$. Thus, we can find that constraints (\ref{C10_UAV_TRANSMIT POWER}) and (\ref{C11_UAV_distance_secure}) are involved in all the ABSs. The corresponding subproblem (\ref{modified_SP_1}) is then reformulated as:
\begin{subequations}
\label{modified_SP_2}
\begin{align}
\underset{\boldsymbol{p}_u, \boldsymbol{\omega}_u, \boldsymbol{d}_u, \boldsymbol{o}_u}{\text{max}} \quad & F(\boldsymbol{\tilde{a}}_u,\boldsymbol{p}_u, \boldsymbol{d}_u; \eta_u), \\
\text { s.t. } \quad & 0 \leq a_{u,m_l} \omega_{u,m_l}(n) \leq p_{\max}, \quad \forall u,m_l,\\
& \sum_{i, j \in \mathcal{U}, i \neq j} f(\boldsymbol{\Theta}_u)^{lb} \geq d_{\mathrm{th}},\\   
& (\ref{C1_UAV}) - (\ref{C9_UAV_DATA_DEMAND_SATISFACTION}),\\
& (\ref{SP_UAV_C1}) - (\ref{SP_UAV_C5}),\\
& (\ref{equality_constraint_pow_1}) - (\ref{equality_constraint_y_1}).
\end{align}
\end{subequations}
\begin{figure*}[b] \centering
\noindent\rule{\textwidth}{0.4pt}\\
%\begin{fleqn}
\small
\begin{equation}  \label{Augmented_lagrangian_function_UAV}
\begin{aligned}[b]
&\mathcal{L}=F(\boldsymbol{\tilde{a}}_u,\boldsymbol{p}_u, \boldsymbol{d}_u; \eta_u)+\sum_{u=1}^{U}  \left( \phi_{u}\left(\boldsymbol{p}_{u}-\boldsymbol{\omega}_{u}\right)+\Pi_{u}\left(x_{u}-\nu_{u}\right)+\Xi_{u}\left(y_{u}-o_{u}\right)\right) +\frac{\rho}{2} \sum_{u=1}^{U} \left ( \left\|\boldsymbol{p}_{u}-\boldsymbol{\omega}_{u}\right\|_{2}^{2} + \left\|x_{u}-\nu_{u}\right\|_{2}^{2} +\left\|y_{u}-o_{u}\right\|_{2}^{2} \right ),
\end{aligned}
\end{equation}
%\end{fleqn}
\small
\begin{equation}   \label{objective_UAV_global}
\begin{aligned}
F(\boldsymbol{\tilde{a}}_u,\boldsymbol{p}_u, \boldsymbol{x}_u, \boldsymbol{y}_u; \eta_u)+\sum_{u=1}^{U}\left( \phi_{u}(\boldsymbol{\tilde{p}}_{u}-\boldsymbol{\omega}_{u})+ \Pi_{u}(\tilde{x}_{u}-\nu_{u})+\Xi_{u}(\tilde{y}_{u}-o_{u}) \right)+\frac{\rho}{2} \sum_{u=1}^{U} \left ( \left\|\boldsymbol{\tilde{p}}_{u}-\boldsymbol{\omega}_{u}\right\|_{2}^{2} + \left\|\tilde{x}_{u}-\nu_{u}\right\|_{2}^{2} + \left\|\tilde{y}_{u}-o_{u}\right\|_{2}^{2}\right ),
\end{aligned}    
\end{equation}
\small
\begin{equation}   \label{Objective_local_uav}
\begin{aligned}[b]   
F(\boldsymbol{\tilde{a}}_u,\boldsymbol{p}_{u}, x_u, y_u; \eta_u)+\phi_{u}(\boldsymbol{p}_{u}-\boldsymbol{\tilde{\omega}}_{u})  +\Pi_{u}(x_{u}-\tilde{\nu}_{u})++\Xi_{u}(y_{u}-\tilde{o}_{u})   +\frac{\rho}{2}\left ( \left\|\boldsymbol{p}_{u}-\boldsymbol{\tilde{\omega}}_{u}\right\|_{2}^{2} + \left\|x_{u}-\tilde{\nu}_{u}\right\|_{2}^{2} + \left\|y_{u}-\tilde{o}_{u}\right\|_{2}^{2}\right ),
\end{aligned}
\end{equation}
\end{figure*}
The problem's augmented Lagrangian function is given by (\ref{Augmented_lagrangian_function_UAV}). We consider that the global copy variables $\boldsymbol{\omega}_{u}$ and $\boldsymbol{\Theta}_{u}$ are managed by the central controller, and the variables $\boldsymbol{p}_{u}$ and $\boldsymbol{d}_{u}$ are processed locally by the ABSs. Based on the above analysis, the global consensus problem for finding global variables $\boldsymbol{\omega}_{u}$ and $\boldsymbol{\Theta}_{u}$ is formulated as follows:
%\begin{fleqn}
\begin{subequations}  \label{Global_problem_ADMM_1_UAV}
\begin{align} 
%\begin{aligned}[b]
\min _{\boldsymbol{\omega}_{u}, \boldsymbol{\Theta}_{u} } \quad & (\ref{objective_UAV_global}), \\
\text {s.t.} \quad & 0 \leq a_{u,m_l} \omega_{u,m_l}(n) \leq p_{\max}, \quad \forall u,m_l,\\
& \sum_{i, j \in \mathcal{U}, i \neq j} f(\boldsymbol{\Theta_u})^{lb} \geq d_{\mathrm{th}},
\end{align}
%\end{aligned}
\end{subequations}
%\end{fleqn}
\begin{algorithm}[t]
\caption{ADMM Distributed Algorithm for Subproblem}
\label{Algorithm_3_ADMM}
\begin{algorithmic}[1]
\State \textbf{Input:} Initialize variables $t$, $\boldsymbol{\phi}$, $\boldsymbol{\Pi}$, $\boldsymbol{\rho}$,
%\State \textbf{Output:} optimal solution $\boldsymbol{a}_u^*$, $\boldsymbol{p}_u^*$, and $\boldsymbol{d}_n^*$
\While {the criterion to stop is not met}
\State \textbf{Central Controller Update}
\State \textbf{continue}
\State \indent wait
\State \textbf{until} obtained updated $\phi_u$, $\Pi_u$, $\Xi_u$, $\boldsymbol{p}_u$, $\boldsymbol{d}_u$ from all \indent ABSs
\State \indent step 1: Solve problem (\ref{Global_problem_ADMM_1_UAV}) and find the optimal \indent \hspace{4 mm} $\boldsymbol{\tilde{\omega}}_{u}$ and $\boldsymbol{\tilde{\Theta}}_{u}$
\State \indent step 2: Send these $\boldsymbol{\tilde{\omega}}_{u}$ and $\boldsymbol{\tilde{\Theta}}_{u}$ to all ABSs
\State \indent step 3: Update the variable t = t+1
\\  \hrulefill
\State \textbf{ABSs Updates}
\State \textbf{continue}
\State \indent wait
\State \textbf{until} from the central controller, obtained updated $\boldsymbol{\tilde{\omega}}_{u}$ and \indent$\boldsymbol{\tilde{\Theta}}_{u}$ 
\State \indent step 1: Solve (\ref{local_problem_ADMM_1_UAV}), and find the optimal solution \indent \hspace{4 mm} $\boldsymbol{\tilde{p}}$ and $\boldsymbol{\tilde{d}}$
\State \indent step 2: All fix valued constraints on dual variables \indent \hspace{4 mm} update: 
\begin{equation*}
\begin{aligned}
& \phi_{u}[t+1]=\phi_{u}[t]+\rho\left(\tilde{\omega}_{u}-\tilde{p}_{u}\right)\\
& \Pi_{u}[t+1]=\Pi_{u}[t]+\rho (\tilde{\nu}_{u}-\tilde{x}_{u})\\
& \Xi_{u}[t+1]=\Xi_{u}[t]+\rho (\tilde{o}_{u}-\tilde{y}_{u})
\end{aligned}
\end{equation*}
\State \indent step 3: Send updated $\boldsymbol{\tilde{p}}$, $\boldsymbol{\tilde{d}}$, $\phi_{u}[t+1]$, $\Pi_{u}[t+1]$ and
\indent \hspace{4 mm} $\Xi_{u}[t+1]$ to the central controller for upcoming 
\indent \hspace{4 mm} iteration
\EndWhile
\end{algorithmic}
\end{algorithm}
where $\boldsymbol{\tilde{p}}_{u}$, $\boldsymbol{\tilde{x}}_{u}$ and $\boldsymbol{\tilde{y}}_{u}$ indicate the constant values which can be obtained by ABSs' update. Therefore, to update $\boldsymbol{p}_{u}$, $\boldsymbol{x}_{u}$ and $\boldsymbol{y}_{u}$, we need to solve the following problem at each ABS $u$:
\begin{subequations}  \label{local_problem_ADMM_1_UAV}
\begin{align} 
%\begin{aligned}[b]
\min _{\boldsymbol{p}_{u}, \boldsymbol{d}_{u} } \quad & (\ref{Objective_local_uav}), \\
\text {s.t.} \quad & (\ref{C1_UAV}) - (\ref{C9_UAV_DATA_DEMAND_SATISFACTION}),\\
& (\ref{SP_UAV_C1}) - (\ref{SP_UAV_C4}).
\end{align}
%\end{aligned}
\end{subequations}
%\end{fleqn}
where $\boldsymbol{\tilde{\omega}}_u$, $\tilde{\nu}_u$ and $\tilde{o}_u$ indicate the fixed values which can be obtained by the central controller's update. Therefore, the dual variables $\phi_u$, $\Pi_u$, and $\Xi_{u}$ can be updated at each ABS $u$ by the following equation:
\begin{subequations}
\begin{align}
\phi_{u}[t+1]=\phi_{u}[t]+\rho\left(\tilde{\omega}_{u}-\tilde{p}_{u}\right),\\
\Pi_{u}[t+1]=\Pi_{u}[t]+\rho (\tilde{\nu}_{u}-\tilde{x}_{u}),\\
\Xi_{u}[t+1]=\Xi_{u}[t]+\rho (\tilde{o}_{u}-\tilde{y}_{u}).
\end{align}
\end{subequations}
The summary of this ADMM is depicted in Algorithm \ref{Algorithm_3_ADMM} and the solution process is shown in Fig. \ref{fig:whole_flowchart}.
We have decomposed and provided iterative algorithms for ABS EE (\ref{optimize_UAV_1}). As we discussed the solution algorithm in detail in earlier sections, in the upcoming problems, these same algorithms will be applied directly to the respective problems without any explanation.\\

\subsection{Satellite Energy Efficiency (Sat-EE)}
This section deals with the satellite EE maximization problem (\ref{optimize_SATELLITE_1}) by utilizing the same previous algorithms. Firstly, BD will apply to decompose, and then DA will transform the subproblem objective into a subtractive form. After that, ADMM will solve the subproblem, which is given as follows:
\subsubsection{Bender Decomposition for Sat-EE} First, the loop counter, i.e., $i_s~$=$~1$ is initialize. Then the variables $\boldsymbol{a}_s$ for $U$ ABSs, and $M_h$ HUEs association is initialized with the upper bound $a_{\textrm{UB}}=1$ lower bound $a_{\textrm{LB}}=0$. Moreover, the function $\chi_s$ as an auxiliary variable, representing the objective function of a subproblem within the master problem's objective function, whose unitize value can be set as $\chi_s=10^{-6}$ to avoid an unbounded solution.\\
\indent \textbf{Subproblem:} We can express the dual variable for the ABSs and HUEs association, i.e., $\kappa_{s,u}$ and $\kappa_{s,m_h}$, respectively. Thus, the subproblem can be define as:
\begin{subequations} \label{SP_Satellite}
\begin{align}
\underset{\boldsymbol{p}_s}{\text{max}} \quad &  \tilde{\eta}_s, \\
\text {s.t.} \quad & \tilde{\eta}_s = \frac{R_{s}\left(\tilde{\boldsymbol{a}}_s, \boldsymbol{p}_s\right)}{P_{s}^{+}},\label{SP_Satellite_C1}\\
& \boldsymbol{a}_{s, u} = \boldsymbol{a}_{s, u}^{i_s} : \kappa_{s, u}^{i_s}, \quad u \in \mathcal{U}, \label{SP_Satellite_C2}  \\
& \boldsymbol{a}_{s,m_h} = \boldsymbol{a}_{s,m_h}^{i_s}: \kappa_{s,m_h}^{i_s}, \quad  \forall m_h \in \mathcal{M}_h, \label{SP_Satellite_C3}\\
& (\ref{C18_SATELLITE_H_UE_DATARATE}), (\ref{C20_SATELLITE_UAV_TRANSMIT_POWER}), (\ref{C21_SATELLITE_H_UE_TRANSMIT_POWER}) \label{SP_Satellite_C5}.
\end{align}
\end{subequations}
\indent \textbf{Convergence Analysis and Bounds:} The objective function at iteration $\psi$ provides the upper bound, which is stated as:
\begin{equation}
\begin{aligned}
\label{upper_bound}
\eta_{\mathrm{UB}}^{{i_s}}=\frac{\tilde{R}_{s}\left(\boldsymbol{a}_s^{{i_s}}, \boldsymbol{p}_s^{{i_s}} \right)}{\tilde{P}_{s}\left(\boldsymbol{a}_s^{{i_s}}, \boldsymbol{p}_s^{{i_s}}\right)^+}.
\end{aligned}
\end{equation}
The lower bound can be define as follows:
\begin{equation}
\label{lower_bound}
\eta_{\mathrm{LB}}^{i_s} = \chi_s^{i_s}.	
\end{equation}
Thus, the stopping criterion can be stated as:
\begin{equation}
\left\{\begin{array}{ll}
\eta_{\mathrm{UB}}^{{i_s}}-\eta_{\mathrm{LB}}^{{i_s}} \leq \epsilon, & \textbf {stop}, \\
\text {otherwise}, & \textbf {continue},
\end{array}\right.
\end{equation}
\indent \textbf{Master Problem:} The loop counter updates as ${i_s} = {i_s}+1$, and after that, the solvable problem becomes as follows:
\begin{subequations} \label{MP_Satellite}
\begin{align}
\underset{\boldsymbol{a}_s, {\chi}_s}{\text{max}} \quad 	& {\chi}_s, \\
\text {s.t.} 	\quad								  	    & (\ref{C14_SATELLITE_HIGH_UE_ASSOCIATION}), (\ref{C15_SATELLITE_UAV_BACKHAUL_ASSOCIATION}),\\
& {\chi}_s \leq \eta_{\mathrm{UB}}^{{i_s}}+\sum_{u=1}^{U} \kappa_{s,u}^{{i_s}}\left(\boldsymbol{a}_{s,u}-\boldsymbol{a}_{s,u}^{{i_s}}\right)  \nonumber \\
& +\sum_{m_h=1}^{M_h} \kappa_{s,m_h}^{{i_s}}\left(\boldsymbol{a}_{s,m_h}-\boldsymbol{a}_{s,m_h}^{{i_s}}\right), \label{bender_cut}\\
& \chi_s \geq \chi^{\text {down}},
\end{align}
\end{subequations}
\subsubsection{Dinkelbach Algorithm for Sat-EE} The objective function in the satellite subproblem can be transformed as follows:
\begin{equation}
F(\boldsymbol{a}_s,\boldsymbol{p}_s; \eta_s) = R_{s}\left(\boldsymbol{\tilde{a}}_s, \boldsymbol{p}_s\right)-\eta_s P_{s}^{+}.
\end{equation}
Thus, the equivalent optimization problem in subtractive form is reformulated as:
\begin{equation}
\label{modified_SP_satellite_1}
\begin{aligned}
\underset{\boldsymbol{p}_s}{\text{max}} \quad & F(\boldsymbol{\tilde{a}}_s,\boldsymbol{p}_s; \eta_s),\\
\text {s.t.} \quad & (\ref{C18_SATELLITE_H_UE_DATARATE}),(\ref{C20_SATELLITE_UAV_TRANSMIT_POWER}), (\ref{C21_SATELLITE_H_UE_TRANSMIT_POWER}).
\end{aligned}
\end{equation}
\subsubsection{ADMM for Sat-EE} In this subproblem problem, we introduce an auxiliary variable $\omega_s$ as a global copy, which implies that a new equality constraint is applied to the subproblem (\ref{modified_SP_satellite_1}), which can be given as:
\begin{equation}
\begin{aligned}
\boldsymbol{p}_{s} = \boldsymbol{\omega}_{s}.  %\quad \forall u \in \mathcal{U}.
\end{aligned}
\end{equation}
We can find that constraints (\ref{C20_SATELLITE_UAV_TRANSMIT_POWER}) and (\ref{C21_SATELLITE_H_UE_TRANSMIT_POWER}) are involved in all the satellite's associated nodes. The corresponding subproblem (\ref{modified_SP_1}) is then reformulated as:
\begin{subequations}
\label{modified_SP_satellite_2}
\begin{align}
\underset{\boldsymbol{p}_s, \boldsymbol{\omega}_s}{\text{max}} \quad & F(\boldsymbol{\tilde{a}}_s,\boldsymbol{p}_s; \eta_s), \\
\text {s.t.} \quad & 0 \leq a_{s,u} \omega_{s,u}(n) \leq p_{\max}, \quad \forall u,\\
& 0 \leq a_{s,m_h} \omega_{s,m_h}(n) \leq p_{\max}, \quad \forall m_h,\\
& (\ref{C18_SATELLITE_H_UE_DATARATE}).
\end{align}
\end{subequations}
The problem's augmented Lagrangian function can be define as follows:
\begin{fleqn}
\begin{equation}  \label{Augmented_lagrangian_function}
\begin{aligned}[b]
& \mathcal{L}=F(\boldsymbol{\tilde{a}}_s,\boldsymbol{p}_s; \eta_s)+\left( \phi_{s}\left(\boldsymbol{p}_{s}-\boldsymbol{\omega}_{s}\right)\right) + \frac{\rho}{2} \left (\left\|\boldsymbol{p}_{s}-\boldsymbol{\omega}_{s}\right\|_{2}^{2}\right )
\end{aligned}
\end{equation}
\end{fleqn}
We consider that the global copy variables $\boldsymbol{\omega}_{s}$ are managed by the central controller, and the variables $\boldsymbol{p}_{s}$ are processed locally by the satellite. Based on the above analysis, the global consensus problem for finding global variables $\boldsymbol{\omega}_{s}$ is formulated as follows:
%\begin{fleqn}
\begin{subequations}  \label{Global_problem_satellite_1}
\begin{align} 
%\begin{aligned}[b]
\min _{\boldsymbol{\omega}_{s} } \quad & F(\boldsymbol{\tilde{a}}_s,\boldsymbol{p}_s; \eta_s)+ \left( \phi_{s}(\boldsymbol{\tilde{p}}_{s}-\boldsymbol{\omega}_{s}) \right)+\frac{\rho}{2} \left ( \left\|\boldsymbol{\tilde{p}}_{s}-\boldsymbol{\omega}_{s}\right\|_{2}^{2} \right ), \\
\text {s.t.} \quad & 0 \leq a_{s,u} \omega_{s, u}(n) \leq p_{\max}, \quad \forall u,\\
\quad & 0 \leq a_{s,m_h} \omega_{s, m_h}(n) \leq p_{\max}, \quad \forall m_h,        
\end{align}
%\end{aligned}
\end{subequations}
%\end{fleqn}
where $\boldsymbol{\tilde{p}}_{s}$ indicates the constant values which can be obtained by the satellite's update. Therefore, to update $\boldsymbol{p}_{s}$, we need to solve the following problem at satellite:
\begin{subequations}  \label{local_problem_ADMM_1_satellite}
\begin{align} 
%\begin{aligned}[b]
\min _{\boldsymbol{p}_{s} } \quad & F(\boldsymbol{\tilde{a}}_s,\boldsymbol{p}_{s}; \eta_s)+\left( \phi_{s}(\boldsymbol{p}_{s}-\boldsymbol{\tilde{\omega}}_{s})\right)+\frac{\rho}{2}\left ( \left\|\boldsymbol{p}_{s}-\boldsymbol{\tilde{\omega}}_{s}\right\|_{2}^{2} \right ),\\
\text {s.t.} \quad & (\ref{C18_SATELLITE_H_UE_DATARATE}).
\end{align}
%\end{aligned}
\end{subequations}
%\end{fleqn}
where $\boldsymbol{\tilde{\omega}}_s$ indicates the fixed value which can be obtained by the central controller's update. Therefore, the dual variable $\phi_u$ can be updated at each ABS $u$ by the following equation:
\begin{subequations}
\begin{align}
\phi_{s}[t+1]=\phi_{s}[t]+\rho\left(\tilde{\omega}_{s}-\tilde{p}_{s}\right).
\end{align}
\end{subequations}

\subsection{Coastline Base Stations Energy Efficiency (CBSs-EE)}
This section deals with the CBSs EE maximization problem (\ref{optimize_CBS_1}) by utilizing the same previous algorithms. Firstly, BD will apply to decompose, and then DA will transform the subproblem objective into a subtractive form. After that, ADMM will solve the subproblem, which is given as follows:
\subsubsection{Bender Decomposition for CBSs-EE} Firstly, the loop counter, i.e., $i_c~$=$~1$ is initialized. Then the variables $\boldsymbol{a}_c$ for $U$ ABS and $M_h$ HUEs association is initialized with the upper bound $a_{\textrm{UB}}=1$ lower bound $a_{\textrm{LB}}=0$. Moreover, the function $\chi_c$ as an auxiliary variable, representing the objective function of a subproblem within the master problem's objective function, whose unitize value can be set as $\chi_c=10^{-6}$ to avoid an unbounded solution. \\
\indent \textbf{Subproblem:} We can express the dual variable for the ABSs and HUEs association i.e., $\kappa_{c,u}$ and $\kappa_{c,m_h}$ respectively. Thus, the subproblem can be define as:
\begin{subequations} \label{SP_CBS}
\begin{align}
\underset{\boldsymbol{p}_c}{\text{max}} \quad &  \tilde{\eta}_c, \\
\text {s.t.} \quad & \tilde{\eta}_c = \frac{R_{c}\left(\tilde{\boldsymbol{a}}_c, \boldsymbol{p}_c\right)}{P_{c}^{+}},\label{SP_CBS_C1}\\
& \boldsymbol{a}_{c,u} = \boldsymbol{a}_{c,u}^{i_c}:\kappa_{c,u}^{i_c}, \quad \forall c \in \mathcal{C}, \forall u \in \mathcal{U}, \label{SP_CBS_C2}  \\
& \boldsymbol{a}_{c,m_h} = \boldsymbol{a}_{c,m_h}^{i_c} : \kappa_{c,m_h}^{i_c}, \quad \forall c \in \mathcal{C}, \forall m_h \in \mathcal{M}_h, \label{SP_CBS_C3}\\
& (\ref{C19_CBS_H_UE_DATARATE}), (\ref{C22_CBS_UAV_TRANSMIT_POWER}), (\ref{C23_CBS_H_UE_TRANSMIT_POWER}) \label{SP_CBS_C5}.
\end{align}
\end{subequations}
\indent \textbf{Convergence Analysis and Bounds:} The objective function at iteration ${i_c}$ provides the upper bound, which is stated as:
\begin{equation}
\begin{aligned}
\label{upper_bound}
\eta_{\mathrm{UB}}^{{i_c}}=\frac{\tilde{R}_{c}\left(\boldsymbol{a}_c^{{i_c}}, \boldsymbol{p}_c^{{i_c}} \right)}{\tilde{P}_{c}\left(\boldsymbol{a}_c^{{i_c}}, \boldsymbol{p}_c^{{i_c}}\right)^+}.
\end{aligned}
\end{equation}
The lower bound can be define as follows:
\begin{equation}
\label{lower_bound}
\eta_{\mathrm{LB}}^{i_c} = \chi_c^{i_c}.	
\end{equation}
Thus, the stopping criterion can be stated as:
\begin{equation}
\left\{\begin{array}{ll}
\eta_{\mathrm{UB}}^{{i_c}}-\eta_{\mathrm{LB}}^{{i_c}} \leq \epsilon, & \textbf {stop}, \\
\text {otherwise}, & \textbf {continue},
\end{array}\right.
\end{equation}
\indent \textbf{Master Problem:} The loop counter update as ${i_c} = {i_c}+1$, and after that the solvable problem become as follows:
\begin{subequations} \label{MP_CBS}
\begin{align}
\underset{\boldsymbol{a}_c, {\chi}_c}{\text{max}} \quad 	& {\chi}_c, \\
\text {s.t.} 	\quad								  	    & (\ref{C16_CBS_HIGH_UE_ASSOCIATION}),(\ref{C17_CBS_UAV_BACKHAUL_ASSOCIATION})\\
& {\chi}_c \leq \eta_{\mathrm{UB}}^{{i_c}} +\sum_{c=1}^{C}\sum_{u=1}^{U} \kappa_{c,u}^{{i_c}}\left(\boldsymbol{a}_{c,u}-\boldsymbol{a}_{c,u}^{{i_c}}\right)  \nonumber \\
& +\sum_{c=1}^{C}\sum_{m_h=1}^{M_h} \kappa_{c,m_h}^{\psi}\left(\boldsymbol{a}_{c,m_h}-\boldsymbol{a}_{c,m_h}^{{i_c}}\right), \label{bender_cut}\\
& \chi_c \geq \chi^{\text {down}},
\end{align}
\end{subequations}
\subsubsection{Dinkelbach Algorithm for CBSs-EE} The objective function in the satellite subproblem can be transformed as follows:
\begin{equation}
F(\boldsymbol{a}_c,\boldsymbol{p}_c; \eta_c) = R_{c}\left(\boldsymbol{\tilde{a}}_c, \boldsymbol{p}_c\right)-\eta_c P_{c}^{+}.
\end{equation}
Thus, the equivalent optimization problem in subtractive form is reformulated as:
\begin{equation}
\label{modified_SP_CBS_1}
\begin{aligned}
\underset{\boldsymbol{p}_c}{\text{max}} \quad & F(\boldsymbol{\tilde{a}}_c,\boldsymbol{p}_c; \eta_c),\\
\text {s.t.} \quad & (\ref{C19_CBS_H_UE_DATARATE}), (\ref{C22_CBS_UAV_TRANSMIT_POWER}), (\ref{C23_CBS_H_UE_TRANSMIT_POWER}).
\end{aligned}
\end{equation}
\subsubsection{ADMM for CBSs-EE} In this subproblem, we introduce an auxiliary variable $\omega_c$ as a global copy, which implies that a new equality constraint is applied to the subproblem (\ref{modified_SP_CBS_1}), which can be given as:
\begin{subequations}
\begin{align}
\boldsymbol{p}_{c} = \boldsymbol{\omega}_{c}.  \quad \forall c \in \mathcal{C}.
\end{align}
\end{subequations}
We can find that constraints (\ref{C22_CBS_UAV_TRANSMIT_POWER}) and (\ref{C23_CBS_H_UE_TRANSMIT_POWER}) are involved in each CBS $c$. The corresponding subproblem (\ref{modified_SP_CBS_1}) is then reformulated as:
\begin{subequations}
\label{modified_SP_CBS_2}
\begin{align}
\underset{\boldsymbol{p}_c, \boldsymbol{\omega}_c}{\text{max}} \quad &     F(\boldsymbol{\tilde{a}}_c,\boldsymbol{p}_c; \eta_c), \\
\text {s.t.} \quad & 0 \leq a_{c,u} \omega_{c,u}(n) \leq p_{\max}, \quad \forall c, u,\\
& 0 \leq a_{c,m_h} \omega_{c,m_h}(n) \leq p_{\max}, \quad \forall c, m_h,\\
& (\ref{C19_CBS_H_UE_DATARATE}).
\end{align}
\end{subequations}
The problem's augmented Lagrangian function can be define as follows:
\begin{fleqn}
\begin{equation}  \label{Augmented_lagrangian_function_CBS}
\begin{aligned}[b]
& \mathcal{L}=F(\boldsymbol{\tilde{a}}_c,\boldsymbol{p}_c; \eta_c)+\left( \phi_{c}\left(\boldsymbol{p}_{c}-\boldsymbol{\omega}_{c}\right)\right) +\frac{\rho}{2} \left (\left\|\boldsymbol{p}_{c}-\boldsymbol{\omega}_{c}\right\|_{2}^{2}\right )
\end{aligned}
\end{equation}
\end{fleqn}
We consider that the global copy variables $\boldsymbol{\omega}_{c}$ are managed by the central controller, and the variables $\boldsymbol{p}_{c}$ are processed locally by each CBS $c$. Based on the above analysis, the global consensus problem for finding global variables $\boldsymbol{\omega}_{c}$ is formulated as follows:
%\begin{fleqn}
\begin{subequations}  \label{Global_problem_satellite_1}
\begin{align} 
%\begin{aligned}[b]
\min _{\boldsymbol{\omega}_{c} } \quad & F(\boldsymbol{\tilde{a}}_c,\boldsymbol{p}_c; \eta_c)+ \left( \phi_{c}(\boldsymbol{\tilde{p}}_{c}-\boldsymbol{\omega}_{c}) \right)+\frac{\rho}{2} \left ( \left\|\boldsymbol{\tilde{p}}_{c}-\boldsymbol{\omega}_{c}\right\|_{2}^{2} \right ), \\
\text {s.t.} \quad & 0 \leq a_{c,u} \omega_{c, u}(n) \leq p_{\max}, \quad \forall c,u,\\
\quad & 0 \leq a_{c,m_h} \omega_{c, m_h}(n) \leq p_{\max}, \quad \forall c,m_h,
\end{align}
%\end{aligned}
\end{subequations}
%\end{fleqn}
where $\boldsymbol{\tilde{p}}_{c}$ indicates the constant values which can be obtained by satellite' update. Therefore, to update $\boldsymbol{p}_{c}$, we need to solve the following problem at each CBS:
\begin{subequations}  \label{local_problem_ADMM_1_satellite}
\begin{align} 
%\begin{aligned}[b]
\min _{\boldsymbol{p}_{c} } \quad & F(\boldsymbol{\tilde{a}}_c,\boldsymbol{p}_{c}; \eta_c)+\left( \phi_{c}(\boldsymbol{p}_{c}-\boldsymbol{\tilde{\omega}}_{c})\right)+\frac{\rho}{2}\left ( \left\|\boldsymbol{p}_{c}-\boldsymbol{\tilde{\omega}}_{c}\right\|_{2}^{2} \right ), \\
\text {s.t.} \quad & (\ref{C19_CBS_H_UE_DATARATE}),
\end{align}
%\end{aligned}
\end{subequations}
%\end{fleqn}
where $\boldsymbol{\tilde{\omega}}_c$ indicates the fixed value which can be obtained by central controller's update. Therefore, the dual variable $\phi_c$ can be updated at each CBS $c$ by the following equation:
\begin{subequations}
\begin{align}
\phi_{c}[t+1]=\phi_{c}[t]+\rho\left(\tilde{\omega}_{c}-\tilde{p}_{c}\right).
\end{align}
\end{subequations}
In the next part, we examine the operation and complexity of algorithms for the proposed problems.\\

\section{Summary and Complexity Analysis}
\label{algorithm_analysis}
As shown in Fig. \ref{fig:whole_flowchart}, to solve the MINLP problem for the SAS-NTN networks, this framework consists of Bender decomposition, the Dinkelbach algorithm, ADMM, and an optimization solver. Bender's decomposition minimizes the complexity of solving the original MILNP by breaking it down into smaller, independent subproblems. Benders' cuts reduce feasible regions with no optimal solution in each iteration. When using ADMM in a subproblem, it will produce an optimal solution in $O(1/\epsilon^2)$ iterations \cite{lin2016iteration}. 

Furthermore, by analyzing the updates in each iteration, the needed complexity for each iteration may be determined. In the ABS EE scenario, updating the power and coordinates requires $O(U\times M_l)$, where $U$ and $M_l$ represent the ABSs and their associated LUEs, respectively. We utilize the convex solver to find a solution because there is no closed-form solution for calculating these variables. As a result, the complexity of these variable update iterations is determined by the solver and the platform employed. Following that, the global update requires $O(U)$ as a projection function, resulting in linear complexity. Finally, we have a constant complexity specified as $O(3)$ for the update of dual variables. It is worth noting that because the number of ABSs $U$ is so small in comparison to the number of LUEs $M_l$ we may ignore it. As a result, the worst-case complexity for a single loop is $O(U\times M)$. If $\Gamma$ is the maximum number of iterations required to reach a sub-optimal solution, then the total execution time of the algorithm is $\Gamma \times O(M\times K)$. This suggests that by adjusting the parameters of $\epsilon,~U$, and $M_l$ our technique can converge within a certain time limit. This same process of complexity analysis is applicable to satellite EE and CBS EE solutions \cite{ADMM_KAZMI}. \\

\section{Simulation Results and Analysis}  
\label{sim}
We now evaluate the performance of our proposed framework. We investigate one random commerce route in the international seas between five ports for HUE travel that is around $500~$km in length. Each port has one CBS that connects the port and the neighboring region to the users. Similarly, LUEs, i.e., fishermen and private boats, are taken into account in these territories up to a $20~$km region in the sea. However, because of LoS linkages and low-gain antennas, these LUEs rely on ABSs for connectivity. As a result, $10$ ABS are stationed in this location, traversing their predetermined course over the sea route for LUEs. As stated in \cite{zong2019optimal}, we assume that worldwide satellite coverage is accessible over the whole studied period of this network.

We consider a $1000~$km x $1000~$km square region with $50$ HUEs spread randomly and equally for our simulations. In the case of LUEs, we assess their dispersion along the neighboring shoreline, where there are $50$ of them and their distribution is random and uniform in a $20~$km x $20~$km square area. All five CBS under consideration are located near the coast, approximately $500~$km apart. At a height of $200~$km, the satellite is deemed in orbit, and all ABSs are initially released into the aerial field at a height of $30$ meters. All statistical results are averaged over a large number of independent experimental iterations in which the initial locations of the LUEs and HUEs are randomized. All simulation results were conducted using Python. Gurobi \cite{gurobi} is an optimizer that is used to solve all optimization problems. Although the simulation does not cover all conceivable circumstances in real-world networking, the results offer an overview of the utility of our proposed strategy. The remaining main parameters are shown in Table \ref{tab1}. 

\begin{figure*}[t] 
\begin{subfigure}{.33\textwidth}
\centering
\includegraphics[scale=0.4]{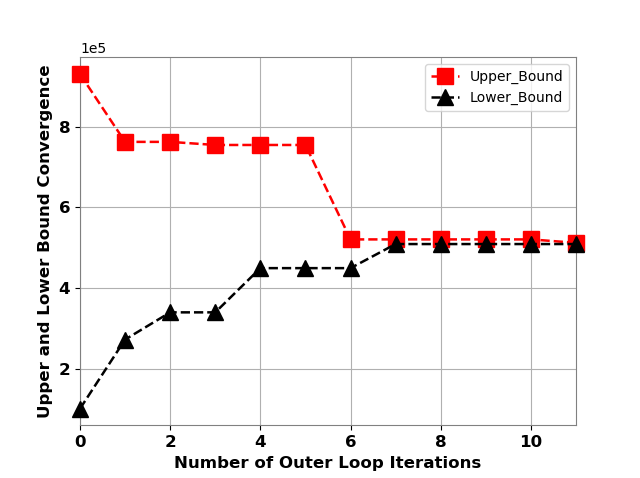} 
\caption{ABS EE convergence.}
\label{convergence_ABS}
\end{subfigure}
\begin{subfigure}{.33\textwidth}
\centering
\includegraphics[scale=0.4]{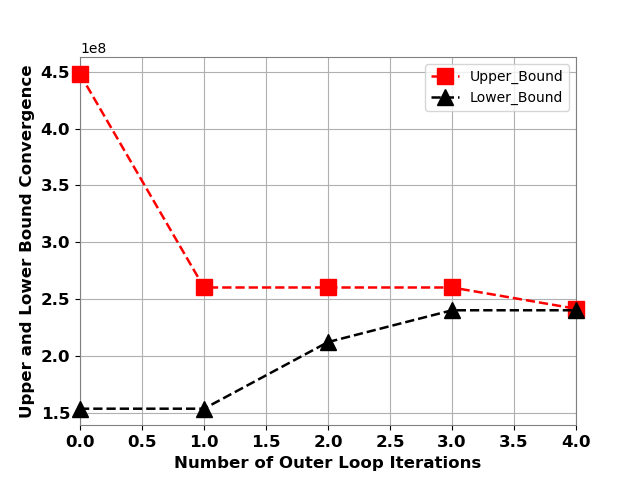} 
\caption{Satellite EE convergence.}
\label{convergence_Sat}
\end{subfigure}
\begin{subfigure}{.33\textwidth}
\centering
\includegraphics[scale=0.4]{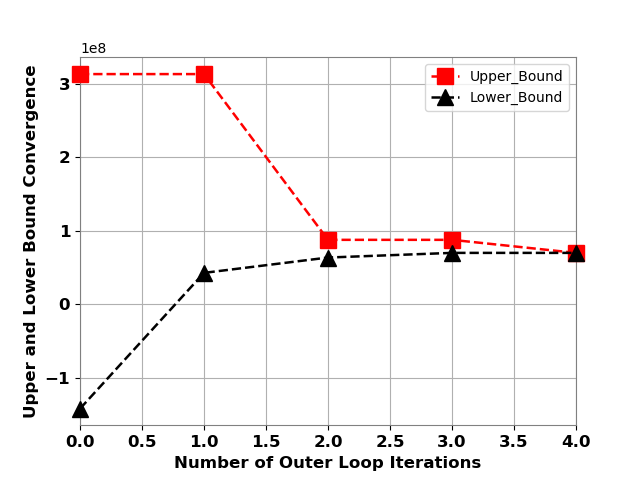}  
\caption{CBS EE convergence.}
\label{convergence_CBS}
\end{subfigure}
\caption{Illustration of the convergence results of three problems i.e., ABS EE, satellite EE, and CBS EE.}
\label{coonvergence}
%\vspace{-0.2in}
\end{figure*}

Fig. \ref{coonvergence} analyzes the convergence of our proposed algorithms for all three problems. The convergence of the ABSs' EE problems can be observed in Fig. \ref{convergence_ABS}. The values of upper bound and lower bound are the optimization goals of the subproblem and master problems, respectively. According to Fig. \ref{coonvergence}, the value of the upper bound is always more than the ideal value, whereas the value of the lower bound is always less than the optimal value. The BD method can converge and approach the suboptimal solution. It can be observed that the BD algorithm for ABS EE converges to a suboptimal solution within 11 iterations. The convergence of satellite EE problems can be found in Fig. \ref{convergence_Sat}. This problem also converged to a suboptimal solution with four iterations. This problem converges more quickly than the ABSs' EE problem due to fewer problem's information sharing with the network controller. Similarly, the convergence of CBS EE is presented in Fig. \ref{convergence_CBS}. This problem also converges rapidly due to less amount of information sharing among network operator and each CBSs. This problem also converges to a suboptimal solution within four iterations.\\
\setlength{\arrayrulewidth}{0.1mm}
\setlength{\tabcolsep}{9pt}
\renewcommand{\arraystretch}{0.8}
\begin{table}[t]
%\centering
\caption{Simulation Parameters} \label{tab1}
\begin{tabular}{|l|l|} \hline
\textbf{Parameters}&\textbf{Values} \\ \hline \hline
CBS radius & $100~$km \\ \hline 
Feasible lower bound & $\chi^{\textrm{down}}~$=$~10^{-6}$\\  \hline
Maximum transmit power & $P^{\textrm{max}}~$=$~33~$dBm  \\ \hline 
Noise power spectral density  & $N_0~$=$~-174~$dBm/Hz   \\ \hline
Carrier frequency  & $f~$=$~30~$GHz \\ \hline 
Satellite and CBS bandwidth & $B_s$,$B_c~$=$~10~$MHz \\ \hline
ABS bandwidth & $B_u~$=$~10~$KHz \\ \hline
%total area  & $100*500~$NM  \\ \hline 
%Satellite area  & $250~$km \\ \hline
%high-end UE speed & $20~$knots \\ \hline 
%low-end UE speed & $25~$knot \\ \hline
%Satellite speed & $7.5~$km/s \\ \hline 
Rician fading channel parameter & $\beta~$=$~1.53$ \\ \hline
HUE Antenna Gain & $G_i~$=$~25~$dBi  \\ \hline
UAV Antenna Gain & $G_u~$=$~25~$dBi  \\ \hline
%Satellite Antenna Gain & $G_s~$=$~30~$dBi \\ \hline
Standard deviation & $\delta_u$,$\delta_{m_h}$, $\delta_c~$=$~0.1$ \\ \hline
reference distance pathloss & ${\omega_{s,u}}$, ${\omega_{c,u}}$, ${\omega_{s, {m_h}}}~$=$~46.4$ \\ \hline
pathloss exponent & ${\zeta_{s,u}}$, ${\zeta_{c,u}}$, ${\zeta_{s, {m_h}}}~$=$~2$ \\ \hline
\end{tabular}
\end{table}
We compared our results with four baseline algorithms, which can be defined as follows:
\begin{itemize}
    \item \textbf{Centralized Algorithm}: This method, which has a complexity of $O(N\times log(N))$, requires a coordinator and demands the entire information as inputs for addressing the defined problem in a centralized way. This scheme can be considered as its results achieve an optimal solution.
    \item \textbf{Greedy Algorithm}: We may use this technique to develop a locally optimal solution that approximates the globally optimal solution at each iteration. In contrast, the greedy algorithm cannot guarantee a globally optimum solution. The algorithm's level of complexity is $O(N^2)$ \cite{ADMM_KAZMI}.
    \item \textbf{Random Algorithm}: This method is distinguished by its degree of unpredictability, which employs uniform random distributions as inputs to achieve excellent performance in terms of average values over all potential input options.
    \item \textbf{Dynamic Programming}: A basic approach that takes into account all of the association and resource allocation pairings and returns suboptimal results. The algorithm's level of complexity is $O(N^2 \times \log N)$.
\end{itemize}

\begin{figure*}[t]
\begin{subfigure}{.33\textwidth}
\centering
\includegraphics[scale=0.4]{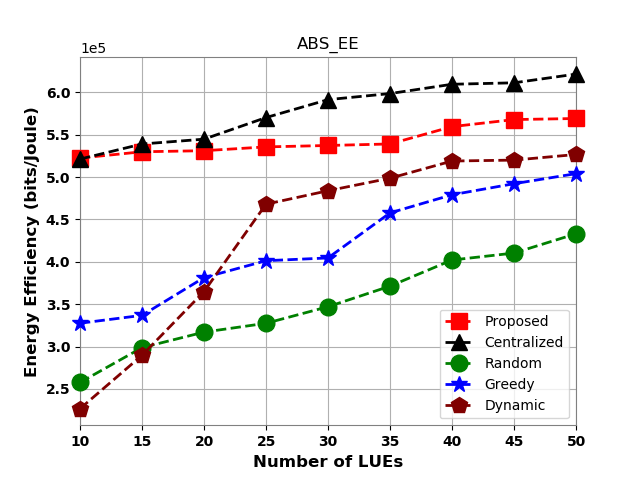} 
\caption{ABS EE vs baselines.}
\label{comparison_ABS}
\end{subfigure}
\begin{subfigure}{.33\textwidth}
\centering
\includegraphics[scale=0.4]{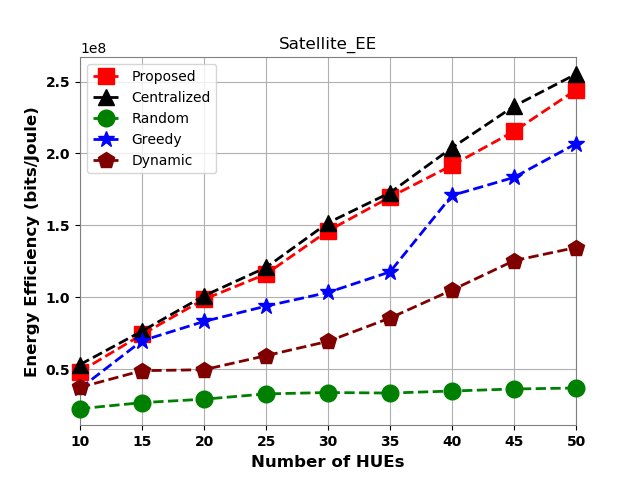} 
\caption{Satellite EE vs baselines.}
\label{comparison_Sat}
\end{subfigure}
\begin{subfigure}{.33\textwidth}
\centering
\includegraphics[scale=0.4]{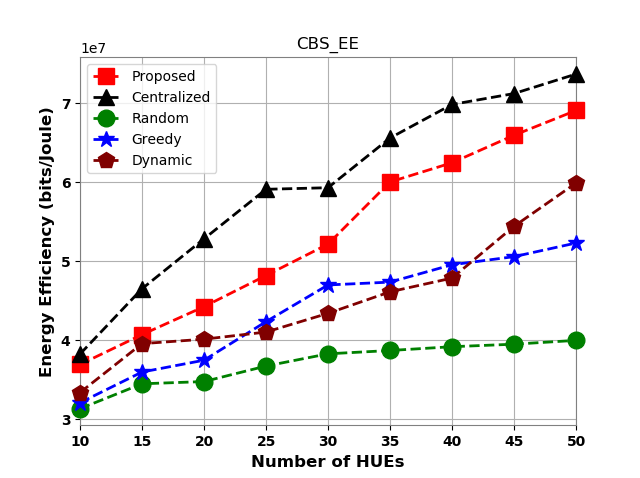}  
\caption{CBS EE vs baselines.}
\label{comparison_CBS}
\end{subfigure}
\caption{Illustration of our proposed algorithms comparison with two baselines for three problems i.e., ABS EE, satellite EE, and CBS EE.}
\label{comparison}
%\vspace{-0.2in}
\end{figure*}

Fig. \ref{comparison_ABS} compares our proposed ABS EE algorithms with the baselines. From this figure, we observe that, when the number of LUEs in the network is set to 10, the proposed algorithm provides the same outcomes as the centralized schemes. Moreover, when the number of LUEs in the network grows, the proposed algorithm produces near-optimal results due to interference and spectrum division in the network. However, the proposed approach outperforms the greedy, random, and dynamic allocation-based algorithms for any number of LUEs. Furthermore, as the number of LUEs associated with the ABSs grows and more bits move through this network, the total energy efficiency of the ABSs also increases, improving network performance. The proposed algorithm for ABS EE achieves up to $27\%$, $12\%$, and $7.7\%$ when compared with random, greedy, and dynamic approaches, respectively, with the number of ABS is set to $10$ and LUEs is set $50$.

\begin{figure*}[t] 
\begin{subfigure}{.33\textwidth}
\raggedleft
\hspace*{-2cm} 
\includegraphics[scale=0.35]{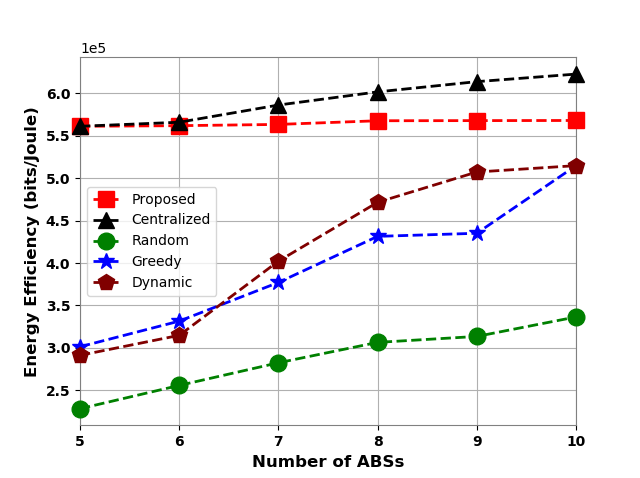}
\caption{ABS EE vs number of deployment}
\label{ABS_variation_for_EE}
\end{subfigure}
\begin{subfigure}{.33\textwidth}
\raggedleft
\hspace*{-2cm} 
\includegraphics[scale=0.35]{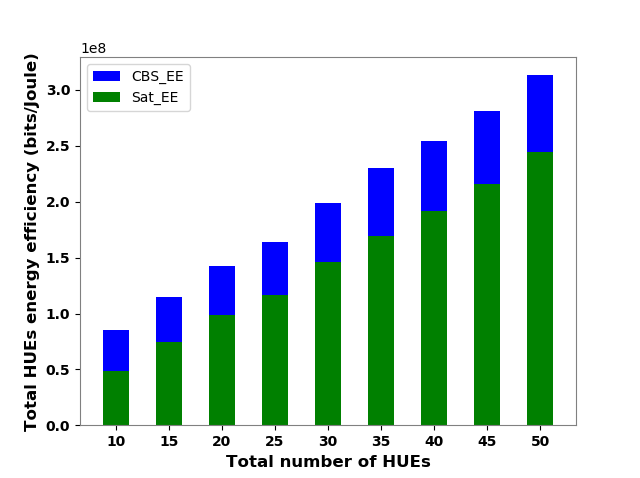}
\caption{EE vs HUEs}
\label{CBS+SAT_HUE}
\end{subfigure}
\begin{subfigure}{.33\textwidth}
\raggedleft
\hspace*{-2cm} 
\includegraphics[scale=0.32]{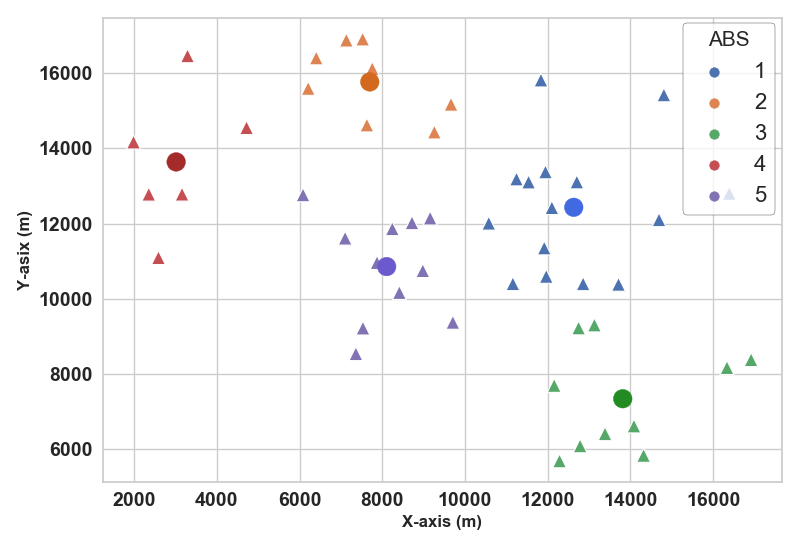}
\caption{ABSs deployment with LUE's association}
\label{ABS_LUE_deployment}
\end{subfigure}
\caption{Illustration of the ABSs' EE, Sat EE, CBS EE, and ABSs' deployment.}
\label{mixed}
%\vspace{-0.2in}
\end{figure*}
Fig. \ref{comparison_Sat} evaluates the EE of the satellite. According to Fig. \ref{comparison_Sat}, our technique achieves near-optimal results for any number of HUEs while the number of ABSs is fixed, which is set at $10$ for satellite-based backhauling. However, under the same network setups, our technique outperforms the greedy, dynamic, and randomized allocation schemes. Furthermore, when the number of HUEs associated with the satellite increases, the network energy efficiency increases due to more data bits traveling across this network. The proposed algorithm for satellite EE achieves up to $16.5\%$ and $57\%$, when compared with greedy and dynamic approaches, respectively, and the number of ABSs is set to $10$ and HUEs is set to $50$.

Fig. \ref{comparison_CBS} shows how our proposed scheme for CBS EE relates to the other four baselines. It is demonstrated that the proposed schemes provide near-optimal results for any number of HUEs with a fixed number of ABSs, which is assumed to be $10$. In this context, the proposed schemes outperform the randomized allocation schemes. Furthermore, as shown in Fig. \ref{comparison_CBS}, when the number of HUEs in the network's surrounding region increases, the EE of the network improves because of an increased amount of bits traveling through the network, thus improving the overall CBSs network performance. The proposed algorithm for CBS EE achieves up to $27\%$ $14.2\%$ and $53.3\%$ when compared with greedy, dynamic, and randomized approaches, respectively, and the number of ABSs is set to $10$ and HUEs is set to $50$.

We show in Fig. \ref{ABS_variation_for_EE} how the number of ABSs deployed in the network affects the performance of ABS EE. We begin by deploying $5$ ABSs in the selected zone and then increase them one by one to assess their impact. It can be shown that as the number of deployed ABSs in the network grows, so does the ABS EE. Furthermore, when the number of ABSs is low, the proposed methodology performs better at first since there is less interference in the system. The total system performs better as the number of ABSs increases gradually, but its relative results with centralized methods are lower due to more power consumption with ABS deployment. However, when compared to greedy, randomized, and dynamic schemes, our proposed algorithm achieves up to $9.8\%$, $51\%$ and $9.83\%$ respectively, with the fixed number of LUEs which is set to $50$. 

In Fig. \ref{CBS+SAT_HUE}, we show how the EE of total HUEs relates to the satellite and CBS's. The satellite has a higher energy efficiency than CBS. This trend has two main reasons: satellites produce their energy from renewable energy sources such as solar energy, which is much less expensive than the running costs of CBS, and according to the system model, a satellite is a more viable network providing source in deep-sea waters than CBS because it can associate multiple HUEs, resulting in better results.

In Fig. \ref{ABS_LUE_deployment}, we demonstrate the ABS deployment in the designated zone. The deployment of 5 ABSs, as well as the LUEs association, are depicted. The ABS positions are denoted by various colored circles. And the ground users are denoted by triangles of the same color as the connecting ABS. The ABS association depends upon the ABS EE maximization by taking into account all the QoS constraints as mentioned in the optimization problem (\ref{optimize_UAV_1}).

\section{Conclusion} \label{conclusion}
In this article, we have studied a maritime wireless communication network that will be used to support future 6G networks. In this network, we designed a novel joint resource allocation of LUEs and HUEs, their association, transmit power control, and the ABSs' deployment problem. We then devised an optimization problem to improve the EE of deployed ABSs, satellites, and CBSs. We have proposed a resource allocation algorithm framework based on joint Benders decomposition, the Dinkelbach algorithm, and the ADMM to handle this problem. This semi-distributed algorithm reduces the processing load on the network's controller while increasing system flexibility. Finally, simulation results show that our proposed method meets the convergence and performance requirements. Future research will investigate the energy and communication efficiency of integrating a larger number of satellites.\\

\begin{appendices}
\section{proof of Lemma 1}
We provide the Taylor approximation of the numerator in (\ref{UAV_objective_first_simple}) with (\ref{taylorexpansion1}). We can define the first-order of the Taylor series as follows:
\renewcommand{\theequation}{A.\arabic{equation}}
\setcounter{equation}{0}
\begin{equation}
f(x_0) + f'(x_0)(x-x_0).
\end{equation}
Let assume $\boldsymbol{d}_{u, \mathrm{local}}$ is local point of $\boldsymbol{d}_{u}$. Now, we can expand Taylor series for this function (\ref{UAV_objective_first_simple}) at point $\boldsymbol{d}_{u, \mathrm{local}}$. Before the Taylor series expansion, let's review a few logarithmic properties. The change of base rule can be given as:
\begin{equation}  \label{taylor_expansion}
\log_e x = \frac{\log_2 x}{\log_2 e},
\end{equation}
it can be modified as:
\begin{equation}
\log_2 x  = ( \log_e x ) ( \log_2 e ) = ( \ln{x} ) ( \log_2 e ),
\end{equation}
therefore, we replace the term $x$ in the natural logarithm with the following term:
\begin{equation}  \label{diff_term}
\ln{ \bigg\{ 1 + \frac{p_{u,m_l}g_0}{(\Omega_{u,m_l} + \sigma^{2}) \Big( \Big\|  \big( \boldsymbol{d}_{u}(n)-\boldsymbol{d}_{m_{l}}(n) \big) \Big\|^{2} \Big) } \bigg \} }. 
\end{equation}
Now, the derivative of natural the logarithmic function can be given as follows:
\begin{equation}
\od{(\ln [f(x)])}{x} = \frac{1}{f(x)} f'(x).
\end{equation}
Now, let's take the derivative of the term given in (\ref{diff_term}) with respect to $\boldsymbol{d}_u$ which is given in (\ref{differentiation}). 
After getting the derivative term, we can put all the terms in (\ref{taylor_expansion}) to get the required expansion term of the objective function, which can be given in (\ref{whole_objective}). We consider that $x_0 = \boldsymbol{d}_{u, local}$ is the local point around which the Taylor series is expressed. Similarly, we put the above-mentioned logarithmic identities in the expansion terms given in (\ref{whole_objective}).
\begin{figure*}[h] \centering
\begin{equation} \label{differentiation}
\begin{aligned}
\frac{\partial }{\partial \boldsymbol{d}_u} \ln{ \bigg\{ 1 + \frac{p_{u,m_l}g_0}{(\Omega_{u,m_l} + \sigma^{2}) \Big( \Big\|  \big( \boldsymbol{d}_{u}(n)-\boldsymbol{d}_{m_{l}}(n) \big) \Big\|^{2} \Big) } \bigg \}} = & \Bigg[ \frac{1}{1+\frac{p_{u,m_l}g_0}{(\Omega_{u,m_l} + \sigma^{2}) \Big( \Big\|  \big( \boldsymbol{d}_{u}(n)-\boldsymbol{d}_{m_{l}}(n) \big) \Big\|^{2} \Big) }} \Bigg] 
\Bigg[ \bigg\{ \frac{{p_{u,m_l}g_0}}{(\Omega_{u,m_l}  + \sigma^{2})} \bigg\} \frac{\partial }{\partial \boldsymbol{d}_u} \Big( \Big\|  \big( \boldsymbol{d}_{u}(n)-\boldsymbol{d}_{m_{l}}(n) \big) \Big\|^{2} \Big)^{-1}          \Bigg] \\ = 
& \Bigg[ \frac{1}{  \frac{{(\Omega_{u,m_l} + \sigma^{2}) \Big( \Big\|  \big( \boldsymbol{d}_{u}(n)-\boldsymbol{d}_{m_{l}}(n) \big) \Big\|^{2} \Big) + {p_{u,m_l}g_0} }}{ (\Omega_{u,m_l} + \sigma^{2}) \Big( \Big\|  \big( \boldsymbol{d}_{u}(n)-\boldsymbol{d}_{m_{l}}(n) \big) \Big\|^{2} \Big) }  }        \Bigg]  
\Bigg[ \bigg\{ \frac{{p_{u,m_l}g_0}}{(\Omega_{u,m_l}  + \sigma^{2})} \bigg\} (-1)  \Big( \Big\|  \big( \boldsymbol{d}_{u}(n)-\boldsymbol{d}_{m_{l}}(n) \big) \Big\|^{2} \Big)^{-2}  \Bigg] \\ =
& \Bigg[ \frac{ -{p_{u,m_l}g_0}}{   \frac{{ \Big\{ (\Omega_{u,m_l} + \sigma^{2}) \Big( \Big\|  \big( \boldsymbol{d}_{u}(n)-\boldsymbol{d}_{m_{l}}(n) \big) \Big\|^{2} \Big) + {p_{u,m_l}g_0}  \Big\}
\Big\{ (\Omega_{u,m_l}  + \sigma^{2})  \Big( \Big\|  \big( \boldsymbol{d}_{u}(n)-\boldsymbol{d}_{m_{l}}(n) \big) \Big\|^{2} \Big)^{2}   \Big\}
}}{ (\Omega_{u,m_l} + \sigma^{2}) \Big( \Big\|  \big( \boldsymbol{d}_{u}(n)-\boldsymbol{d}_{m_{l}}(n) \big) \Big\|^{2} \Big) }} \Bigg]  \\ =
& \Bigg[ \frac{ -{p_{u,m_l}g_0} } { \Big\{ (\Omega_{u,m_l} + \sigma^{2}) \Big( \Big\|  \big( \boldsymbol{d}_{u}(n)-\boldsymbol{d}_{m_{l}}(n) \big) \Big\|^{2} \Big) + {p_{u,m_l}g_0}  \Big\} \Big\{  \Big\|  \big( \boldsymbol{d}_{u}(n)-\boldsymbol{d}_{m_{l}}(n) \big) \Big\|^{2} \Big\} } \Bigg]
\end{aligned}
\end{equation}
\begin{equation}  \label{whole_objective}
\small
\begin{aligned}
\left[ \log_2 \bigg\{ 1 + \frac{p_{u,m_l}g_0}{(\Omega_{u,m_l} + \sigma^{2}) \Big( \Big\|  \big( \boldsymbol{d}_{u, \mathrm{local}}(n)-\boldsymbol{d}_{m_{l}}(n) \big) \Big\|^{2} \Big) } \bigg \}-
\frac{ p_{u,m_l}g_0 \Big\{ \Big\| \big( \boldsymbol{d}_{u}(n)-\boldsymbol{d}_{m_{l}}(n) \big) \Big\|^{2} - \Big\| \big( \boldsymbol{d}_{u, \mathrm{local}}(n)-\boldsymbol{d}_{m_{l}}(n) \big) \Big\|^{2} \Big\} \log_2 e}{ \Big\{ \Big\| \big( \boldsymbol{d}_{u, \mathrm{local}}(n)-\boldsymbol{d}_{m_{l}}(n) \big) \Big\|^{2} \Big\} \Big\{p_{u,m_l}g_0 +  (\Omega_{u,m_l} + \sigma^{2}) \Big( \Big\| \big( \boldsymbol{d}_{u, \mathrm{local}}(n)-\boldsymbol{d}_{m_{l}}(n) \big) \Big\|^{2} \Big) \Big\} }  \right]
\end{aligned}
\end{equation}
\noindent\rule{\textwidth}{0.4pt}
\end{figure*}
\section{proof of Lemma 2}
\renewcommand{\theequation}{B.\arabic{equation}}
\setcounter{equation}{0}
We applied the Taylor series expansion to the quadratic safe distance constraint, which makes it linearize and can be solved with a standard solver. Here, we consider the local point $\boldsymbol{d}_{u, local}$ for each ABS around which the Taylor series applies. The first-order Taylor series can be expressed as:
\begin{equation} \label{Taylorexpansion_distance}
f(x_0) + f'(x_0)(x-x_0).
\end{equation}
Here, function can be represented as:
\begin{equation} \label{function_distance1}
f(\boldsymbol{d}_u) = \| \boldsymbol{d}_{u1} - \boldsymbol{d}_{u2} \|^2.
\end{equation}
We can simply the norm function as follows:
\begin{equation} \label{function_distance2}
f(\boldsymbol{d}_u) = \big \langle \boldsymbol{d}_{u1} - \boldsymbol{d}_{u2}, \boldsymbol{d}_{u1} - \boldsymbol{d}_{u2}   \big \rangle 
\end{equation}
Let's take the first-order derivative of function (\ref{function_distance2}) according to Leibniz formula which can be expressed as:
\begin{equation}
(uv)' = u'v + uv'.
\end{equation}
Lets put our function into the above the equation:
\begin{equation}
f(\boldsymbol{d}_u)' = \boldsymbol{d}_{u1} - \boldsymbol{d}_{u2} + \boldsymbol{d}_{u1} - \boldsymbol{d}_{u2},
\end{equation}
which can be simplify as follows:
\begin{equation}
f(\boldsymbol{d}_u)' = 2(\boldsymbol{d}_{u1} - \boldsymbol{d}_{u2}),
\end{equation}
Lets combine all the terms and put in first order Taylor series as given in (\ref{Taylorexpansion_distance} ):
\begin{equation}
\| \boldsymbol{d}_{u1, \mathrm{local}} - \boldsymbol{d}_{u2, \mathrm{local}} \|^2 + 2(\boldsymbol{d}_{u1} - \boldsymbol{d}_{u2})\cdot(\boldsymbol{d}_{u1, \mathrm{local}} - \boldsymbol{d}_{u2, \mathrm{local}}),
\end{equation}
which can be modified as follows by applying the dot product property of transposition:
\begin{equation}
\| \boldsymbol{d}_{u1, \mathrm{local}} - \boldsymbol{d}_{u2, \mathrm{local}} \|^2 + 2(\boldsymbol{d}_{u1, \mathrm{local}} - \boldsymbol{d}_{u2, \mathrm{local}})\cdot(\boldsymbol{d}_{u1} - \boldsymbol{d}_{u2})^T.
\end{equation}
So this is the simplified first-order Taylor expansion of the safe distance quadratic constraint.\\
\end{appendices}
	
\bibliographystyle{IEEEtran}
\bibliography{ee_ref}
\end{document}